\documentclass[%
 reprint,
superscriptaddress,
nofootinbib,
 amsmath,amssymb,
 aps,
]{revtex4-2}
\usepackage{graphicx}
\usepackage{dcolumn}
\usepackage{bm}
\usepackage[version=4]{mhchem} 
\usepackage{subfig}
\usepackage{float}
\usepackage{caption}
\usepackage[colorlinks,
            linkcolor=red,
            anchorcolor=blue,
            citecolor=blue
            ]{hyperref}

\graphicspath{ {figure/} }
\begin{document}

\title{Reanalysis of the Systematic Uncertainties in Cosmic-Ray Antiproton Flux}

\author{Xing-Jian Lv}
\email{lvxj@ihep.ac.cn}
 \affiliation{%
 Key Laboratory of Particle Astrophysics, Institute of High Energy Physics, Chinese Academy of Sciences, Beijing 100049, China}
\affiliation{
 University of Chinese Academy of Sciences, Beijing 100049, China 
}%
 \author{Xiao-Jun Bi}
 \email{bixj@ihep.ac.cn}
\affiliation{%
 Key Laboratory of Particle Astrophysics, Institute of High Energy Physics, Chinese Academy of Sciences, Beijing 100049, China}
\affiliation{
 University of Chinese Academy of Sciences, Beijing 100049, China 
}%
\author{Kun Fang}
\email{fangkun@ihep.ac.cn}
\affiliation{%
 Key Laboratory of Particle Astrophysics, Institute of High Energy Physics, Chinese Academy of Sciences, Beijing 100049, China}
 \author{Peng-Fei Yin}
\email{yinpf@ihep.ac.cn}
\affiliation{%
 Key Laboratory of Particle Astrophysics, Institute of High Energy Physics, Chinese Academy of Sciences, Beijing 100049, China}
\author{Meng-Jie Zhao}
\email{zhaomj@ihep.ac.cn}
 \affiliation{%
 Key Laboratory of Particle Astrophysics, Institute of High Energy Physics, Chinese Academy of Sciences, Beijing 100049, China}
\affiliation{
 University of Chinese Academy of Sciences, Beijing 100049, China 
}%



\date{\today}

\begin{abstract}
Recent studies on cosmic rays (CRs) have reported the possibility of an excess in the antiproton flux around $10-20$ GeV. However, the associated systematic uncertainties have impeded the interpretation of these findings. In this study, we conduct a global Bayesian analysis to constrain the propagation parameters and evaluate the CR antiproton spectrum, while comprehensively accounting for uncertainties associated with interstellar CR propagation, production cross sections for antiprotons and other secondaries, and the charge and energy dependent effects of solar modulation. We establish that the most recent AMS-02 $\bar{p}$ spectrum is in agreement with a pure secondary origin. Based on this, we establish upper limits on dark matter (DM) annihilation. We also determine that the AMS-02 data favors the empirical hadronic interaction models over phenomenological ones. 
Finally, we find that the latest AMS-02 antiproton data from 2011 to 2018 disfavors the antiproton excess at $\mathcal{O}$(10) GeV and the corresponding DM interpretation that can simultaneously account for the Galactic Center excess in the gamma-ray observation.
\end{abstract}
\maketitle


\section{\label{sec:level1}INTRODUCTION}
Antimatter investigations in cosmic rays (CRs) have critical implications for the fields of astrophysics and dark matter (DM) indirect detection. Over the past few years, significant advancements have been made in the measurement of antimatter particles in CRs. Of particular note is the Alpha Magnetic Spectrometer (AMS-02)~\cite{AMS02AlphaMagnetic}, which was launched in 2011 and has yielded extremely precise CR measurements. The characteristics of CR antimatter particles have been extensively analyzed quantitatively in the literature~\cite{Blum:2017iwq}, drawing on the findings of previous experiments such as PAMELA~\cite{PAMELAExperimentWeb} and Fermi-LAT~\cite{FermiGammaraySpace}, in conjunction with those from AMS-02.

Several recent studies have reported an overabundance of antiprotons in the flux at energies between 10 and 20 GeV, which cannot be explained by their production as secondary particles alone~\cite{Cui:2016ppb, Cuoco:2016eej, Cui:2018klo, Cuoco:2019kuu, Cholis:2019ejx, Zhu:2022tpr}. These studies proposed that the excess might be attributed to the annihilation or decay of some weakly interacting massive particles (WIMPs), which are candidates of DM. The proposed WIMPs have a mass in the range of $50\sim100$ GeV and a thermally-averaged annihilation cross section $\langle\sigma v\rangle$ near the so-called nature value of $\sim3\times 10^{-26}$ cm$^3$s$^{-1}$, and predominantly annihilate to some hadronic final states. Additionally, several authors suggested that these WIMPs can explain the Galactic Center excess (GCE) observed in gamma-ray emissions by Fermi-LAT~\cite{DiMauro:2021qcf, Cuoco:2017rxb, DiMauro:2021raz}, and are in line with other direct or indirect DM search results~\cite{Leane:2020liq, Hooper:2019xss}. On the other hand, many recent studies also argued that the antiproton flux is consistent with a pure secondary origin~\cite{Boudaud:2019efq, Heisig:2020nse, Reinert:2017aga, caloreAMS02AntiprotonsDark2022, Lin:2016ezz, Lin:2019ljc, Luque:2021ddh}.

At the crux of the present controversy lies the intricacy of accurately predicting the secondary antiproton flux. These antiprotons are generated via the interaction between high energy CRs and the interstellar gas, with their flux being determined by the process of CR propagation and its hadronic interaction with the interstellar gas~\cite{Donato:2001ms}. However, both of these factors have considerable uncertainties. In this study, we present a systematic analysis of the secondary antiproton flux, wherein we deliberately consider the aforementioned sources of uncertainty. Then we can derive constraints on the cross section of DM annihilation based on the predicted $\bar{p}$ flux and the AMS-02 data.

The propagation of CRs within the Galaxy is a complex process that
entails diverse phenomena including diffusion, energy loss, convection, and reacceleration. 
Stable primary-to-secondary ratios, like B/C and B/O, are the main probes for studying CR propagation. 
Recent analysis
utilizing high-precision data provided by the AMS-02 collaboration has indicated
that a modified diffusion-reacceleration model (hereafter referred to as DR2 model) is preferred~\cite{Yuan:2017ozr, Yuan:2018vgk}. However,
the present measurements of CR data are insufficient to pinpoint
the various propagation parameters as the uncertainties involved in secondary production cross sections can bias these parameters~\cite{tomassettiSolarNuclearPhysics2017, Genolini:2018ekk, Weinrich:2020cmw, Korsmeier:2021brc}.
The solar modulation effect on CRs induces further uncertainties in propagation as it significantly affects low-energy CR spectra, while its precise quantification remains challenging~\cite{tomassettiSolarNuclearPhysics2017, Potgieter:2013pdj}.

We employ the numerical tool GALPROP v56\footnote{Current version available at \url{https://galprop.stanford.edu/}}~\cite{Strong:1998pw, Strong:1998fr} to solve the CR propagation equation. The propagation parameters are constrained by the Bayesian analysis, where the latest data on B/C and B/O are used. A Markov Chain Monte Carlo (MCMC) algorithm~\cite{Lewis:2002ah} is applied to explore multidimensional parameter space with high efficiency. The uncertainties in cross sections for Boron production are fully taken into account to eliminate potential bias in the resulting propagation parameters. A charge-sign dependent solar modulation model~\cite{Cholis:2015gna} is adopted to describe the different modulation processes for $p$ and $\bar{p}$.

Another important source of uncertainties in predicting secondary antiproton flux is the $\bar{p}$ production cross section. Fortunately, its uncertainty has been reduced owing to the advancements in collider experiments~\cite{NA49:2012jna, NA61SHINE:2017fne, LHCb:2018ygc}. We adopt the cross sections derived in Ref. ~\cite{Korsmeier:2018gcy} that incorporate the most recent data. The uncertainties in $\bar{p}$ production are fully propagated into the final flux. Additionally, we compare the performance of the latest model with some older models that do not incorporate the latest data, aiming to investigate the potential differences between them.


The paper is organized as follows. In Sec. \ref{sec methology} we detail the propagation model and methodology employed to derive the propagation and source parameters which are the basis of the calculation of the background antiproton flux. Our setup of DM annihilation is also introduced in this section. In Sec. \ref{sec results} we show the resulting secondary antiproton spectrum under various settings and its implications on DM annihilation. Finally, we summarize our findings and provide insightful discussions in Sec. \ref{sec:conclusion}.

\section{\label{sec methology}METHOD}
\subsection{\label{sec:model}CR propagation model}%
Following acceleration within sources, galactic CRs undergo injection and diffusion in the interstellar medium, subsequently experiencing various propagation effects before reaching the Earth. It is commonly assumed that the propagation of CRs takes place within a cylindrical halo with a vertical extent of half height $L$, beyond which CRs are able to escape freely. This process can be described mathematically with a propagation equation~\cite{Strong:2007nh}
\begin{equation}
\begin{aligned}
\frac{\partial \psi}{\partial t}= & Q(\mathbf{x}, p)+\nabla \cdot\left(D_{x x} \nabla \psi-\mathbf{V}_c \psi\right)+\frac{\partial}{\partial p}[p^2 D_{p p} \frac{\partial}{\partial p}(\frac{\psi}{p^2})] \\
& -\frac{\partial}{\partial p}[\dot{p} \psi-\frac{p}{3}(\nabla \cdot \mathbf{V}_c) \psi]-\frac{\psi}{\tau_f}-\frac{\psi}{\tau_r}\;,
\end{aligned}
\end{equation}
where $Q(x, p)$ denotes the CR source term, $\psi = \psi(x, p, t)$ is the CR density per momentum interval, $ \dot{p} \equiv dp/dt$ represents the momentum loss rate, and the time scales $\tau_f$ and $\tau_r$ characterize fragmentation processes and radioactive decays, respectively.

Supernova remnants are widely recognized as the main sources of Galactic CRs, wherein charged particles experience acceleration through shock waves. Consequently, the spatial distribution of the CR sources follows that of supernova remnants~\cite{Trotta:2010mx}. According to the shock acceleration theory~\cite{Achterberg:2001rx}, the injection spectrum of primary CRs follows a power-law distribution that is dependent on the rigidity of CRs, $q\propto R^{-\nu}$. To account for the observed structures in CR spectra, a three-piece broken power law is adopted to describe the rigidity dependence of the injection spectrum~\cite{Ahn:2010gv}.

The spatial diffusion coefficient $D_{xx}$ can be parametrized as~\cite{Maurin:2010zp}
\begin{equation}
D_{x x}=D_0 \beta^\eta\left(R / R_0\right)^\delta
\end{equation}
where $R \equiv pc/Ze$ is the rigidity of the CR particle, $\beta$ is the CR particle velocity in units of the speed of light $c$, and $R_0$ and $D_0$ represent the reference rigidity and normalization parameter, respectively. Although the theoretical prediction for the slope of the diffusion coefficient $\delta$ is expected to be 1/3 for a Kolmogorov spectrum of interstellar turbulence or 1/2 for a Kraichnan cascade, it is generally treated as a free parameter to fit the data. Additionally, the slope of the velocity $\eta$ reflects the possibility that the turbulence dissipation could alter the diffusion coefficient at low velocities~\cite{Yuan:2017ozr}.

The process of reacceleration of CRs caused by collisions with interstellar weak hydrodynamic waves can be characterized by the diffusion in momentum space, which is determined by the momentum diffusion coefficient $D_{pp}$. This coefficient is interrelated with the spatial diffusion coefficient $D_{xx}$ through~\cite{Drury:2016ubm}
\begin{equation}
D_{p p} D_{x x}=\frac{4 p^2 v_A^2}{3 \delta\left(4-\delta^2\right)(4-\delta) \omega}\;,
\end{equation}
where $v_A$ and $\omega$ denote the Alfv\'en velocity and the ratio of magnetohydrodynamic wave energy density to magnetic field energy density, respectively. Since $D_{pp}\propto v^2_A/\omega$, we can set $\omega=1$ without loss of generality. Furthermore, previous research has indicated that convection is not favored by the observed CR spectrum~\cite{Yuan:2018vgk}. Thus, for the purposes of this study, $\mathbf{V}_c$ is set to $0$ throughout the remainder of the article.

The CR spectra are greatly influenced by solar modulation below $\sim20$ GeV. The conventional approach to address this phenomenon involves utilizing the force-field approximation, with the solar modulation potential $\phi$ employed to quantify the associated strength~\cite{Gleeson:1968zza}. However, the use of this approach in situations involving particles of opposite charges is deemed inadequate due to the presence of drift effects~\cite{Potgieter:2013pdj}. In Ref. ~\cite{Cholis:2015gna}, a modified technique was proposed, incorporating a simple amendment to the solar modulation potential's structure to account for this challenge
\begin{equation}
\phi^{ \pm}(t, R)=\phi_0(t)+\phi_1^{ \pm}(t) \mathcal{F}\left(R / R_0\right)\;,
\end{equation}
where $\mathcal{F}=(1+\left(R / R_0\right)^2)/(\beta\left(R / R_0\right)^3)$ and $R_0=0.5$ GV is the reference rigidity. The second term on the right-hand side of the equation accounts for the augmented energy dissipation encountered by particles whose charge polarity is mismatched with the current sheet. For the positive (negative) polarity phase, the values of $\phi^{+}_1$($\phi^{-}_1$) are assigned as 0. 

In line with previous studies ~\cite{Reinert:2017aga, Luque:2021ddh}, we fix the value of $\phi^{+}_{1}$ to 0, while treating the parameters $\phi_{0}$ and $\phi^{-}_{1}$ as time-averaged values during the experimental process, which are obtained by fitting the CR data. We further allow for the possibility that $\phi^{+}_{0}$ and $\phi^{-}_{0}$ may have distinct values, as particles with opposite charges traverse different regions of the solar system~\cite{straussModellingHeliosphericCurrent2012}. The results show that they only differ by $\mathcal{O}(0.1\mathrm{ GV})$.

\subsection{\label{subsec ana}Analysis setup}
We employ a two-step approach, as previously proposed in Ref. ~\cite{Cui:2016ppb, Cui:2018klo}, to compute the antiproton spectrum. The first step entails utilizing an MCMC technique based on Bayesian inference to derive the propagation parameters $D_0$, $\delta$, $v_A$, and $\eta$, and their corresponding covariance matrix from the secondary to primary ratios B/C and B/O. In consideration of the strong dependence of the DM induced signature flux on the propagation halo height $L$ and the conflict between the Be/B and $\ce{^{10}Be}/\ce{^{9}Be}$ data on $L$~\cite{Weinrich:2020ftb, DeLaTorreLuque:2021yfq, Evoli:2019iih}, $L$ is not fitted in this study. Instead, two benchmark values $L=3.44$ and 7.17 Kpc are selected, which are inferred from the combined B/C+$\ce{^{10}Be}/\ce{^{9}Be}$ and combined B/C+$\ce{^{10}Be}/\ce{^{9}Be}$+Be/B fits, respectively~\cite{Zhao:2022bon}. 
Additionally, acknowledging the implications of synchrotron data, which suggest a larger halo as ~\cite{Orlando:2013ysa}, we systematically explore a broader range of halo heights -- specifically, 2, 4, 6, 8, and 10 kpc. This approach is implemented in our determination of the upper limits on the DM annihilation cross section, thereby encompassing the full range of uncertainties associated with this critical parameter.
The B/C and B/O ratios are rather insensitive to the source spectra of their progenitors and the choice of the Fisk potential $\phi$, thus enabling us to maintain a fixed value for the primary injection spectrum and $\phi$ for the sake of simplicity. A preliminary fit yields the values of these parameters, which are tabulated in Table \ref{table injection}.

To incorporate the uncertainties in the Boron production cross section, a data driven method from earlier research is used~\cite{Tomassetti:2015nha, tomassettiSolarNuclearPhysics2017, Zhao:2022bon}, whereby $\sigma_{\mathrm{prod.}}$ of 7 different channels are renormalized to match the available data points through overall rescaling. These central values and dispersions are included as priors during the fit and can be found in table III of Ref. ~\cite{Zhao:2022bon}. Consequently, the uncertainties in the production cross section are naturally propagated into the transportation parameters.

In the second step, we determine the injection spectrum of protons and Helium by performing a combined MCMC fit of the proton, Helium, and antiproton spectra. Additionally, we conduct an ancillary fitting procedure that excluded antiprotons, to ascertain whether the results obtained are consistent. Furthermore, we allow the propagation parameters to vary to account for the possibility of non-universality of CR transportation~\cite{Johannesson:2016rlh, Zhao:2022bon}. Multivariate Gaussian priors on propagation parameters are utilized to incorporate the central values and covariance matrix of the propagation parameters from the first step. 

The production cross section of the antiproton is an essential factor in calculating their spectrum, and we utilize the parametrization proposed in Ref. ~\cite{Winkler:2017xor} and improved in Ref. ~\cite{Korsmeier:2018gcy} (hereafter referred to as Winkler + Korsmeier), which account for the effects of isospin violation, hyperon contribution, and scaling violation. To incorporate the uncertainties in the antiproton production cross section, we introduce an overall rescaling factor $XS_{\bar{p}}$ with a Gaussian prior of $\mathcal{N}(1.0, 0.08)$ to reproduce the uncertainty band shown in Fig. 5 of Ref. ~\cite{Korsmeier:2018gcy}.

\begin{table}[]
\captionsetup{justification=raggedright}
\caption{\label{table injection}Injection spectrum of primary particles (except protons and Helium) and Fisk potential of heavy elements adopted in this study}
\begin{tabular}{c|c|c|c|c|c}
$\nu_1$  & $\nu_2$  & $\nu_3$ & $R_{\mathrm{br1}}$(GV) & $R_{\mathrm{br2}}$(GV) & $\phi$(GV)                \\ \hline
1.23 & 2.38 & 2.24 & 2.1                    & 222.8                  & \multicolumn{1}{c}{0.788}
\end{tabular}
\end{table}

Regarding data selection, we employ the most up-to-date CR data available to us. Specifically, the B/C and B/O data are sourced from AMS-02 observations spanning the 2011-2016 period~\cite{AMS:2021nhj}. We further utilize proton flux data collected by Voyager (2012-2015)~\cite{Cummings:2016pdr}, AMS-02 (2011-2018)~\cite{AMS:2021nhj}, and DAMPE (2016-2018)~\cite{DAMPE:2019gys} to derive the proton injection spectral parameters. For Helium, the AMS-02 (2011-2018)~\cite{AMS:2021nhj} and Voyager (2012-2015)~\cite{Cummings:2016pdr} data are used. Notably, AMS-02 proton and Helium data falling below 6 GV are excluded from the analysis since they are below the antiproton production threshold~\cite{Donato:2001ms}. Finally, the $\bar{p}$ data are taken from AMS-02 observations during the 2011-2018 period~\cite{AMS:2021nhj}.

In this study, the MCMC procedure is implemented using the publicly available codebase, cobaya\footnote{\url{https://cobaya.readthedocs.io/en/latest/}}~\cite{Lewis:2002ah, Lewis:2013hha, Torrado:2020dgo}. The Metropolis-Hastings algorithm is employed to generate samples from the posterior distribution, which is a key step in Bayesian statistical inference. In addition to the MCMC sampler, we utilize the Python package GetDist\footnote{\url{https://getdist.readthedocs.io/en/latest/}}~\cite{Lewis:2019xzd} which offers a suite of analytical tools that facilitate the interpretation of the posterior distribution, including methods for estimating credible intervals and producing confidence contour plots.

In addition to the default setup, we perform several alternative analyses to ensure the robustness of our results. One such approach involved examining the use of different models for antiproton production cross sections. These models can be categorized as either phenomenological or empirical. The former are based on Monte Carlo generators and are often used to model hard-scattering processes in colliders, which are not necessarily well-suited to CR antiproton production dominated by soft production. Conversely, the empirical models are derived from the fits to measurement results with certain analytical parametrization. The model utilized in our default analysis belongs to the second category and was fitted to the latest NA61 experiment~\cite{NA61SHINE:2017fne} in the $pp$ channel and the first ever determination of $p$He data by the LHCb experiment~\cite{LHCb:2018ygc}. We also examine the use of the default hadronic model embedded in GALPROP, which uses the parametrization proposed by Tan \& Ng~\cite{Tan:1983kgh}. In addition, we explore two phenomenological models that are best suited to CR studies according to Ref. ~\cite{Lin:2016ezz}, QSJET II-04m~\cite{Kachelriess:2015wpa} and LHC 1.99~\cite{Pierog:2009zt}, both of which are tuned to low-energy data.

Moreover, we have conducted a quantitative assessment of the influence that the latest antiproton data from AMS-02 2011-2018~\cite{AMS:2021nhj} exerts when compared to the previously employed data of AMS-02 2011-2015~\cite{AMS:2016oqu}, which remains the most widely utilized data, even in contemporary publications. Our analysis reveals that the utilization of the new data manifests a nuanced yet significant impact on the ultimate outcomes, as demonstrated in Sec. \ref{subsec:new data}.

\begin{figure*}
    \begin{center}
        \subfloat{\includegraphics[width=0.45\textwidth]{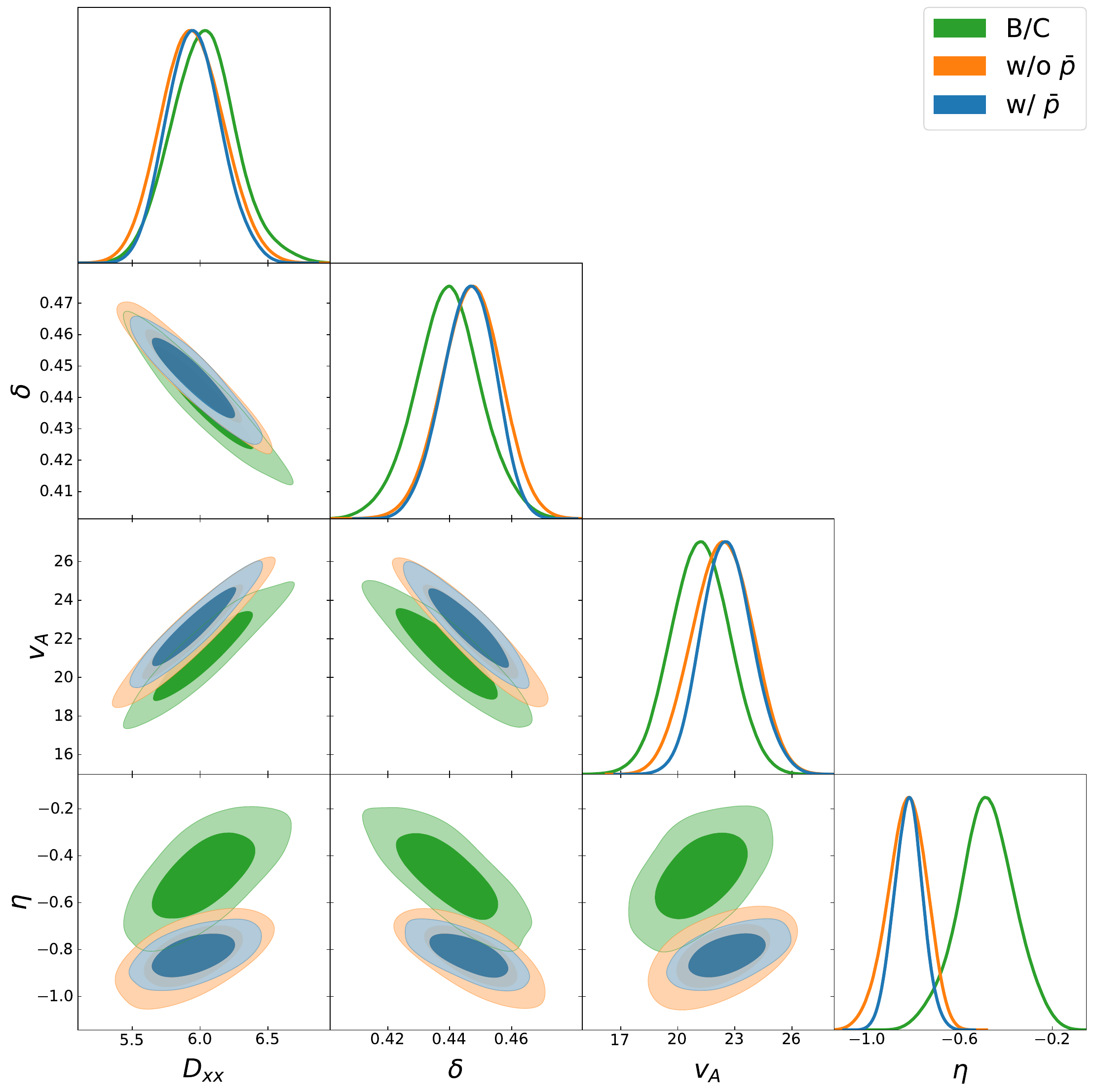}} \hskip 0.03\textwidth
        \subfloat{\includegraphics[width=0.45\textwidth]{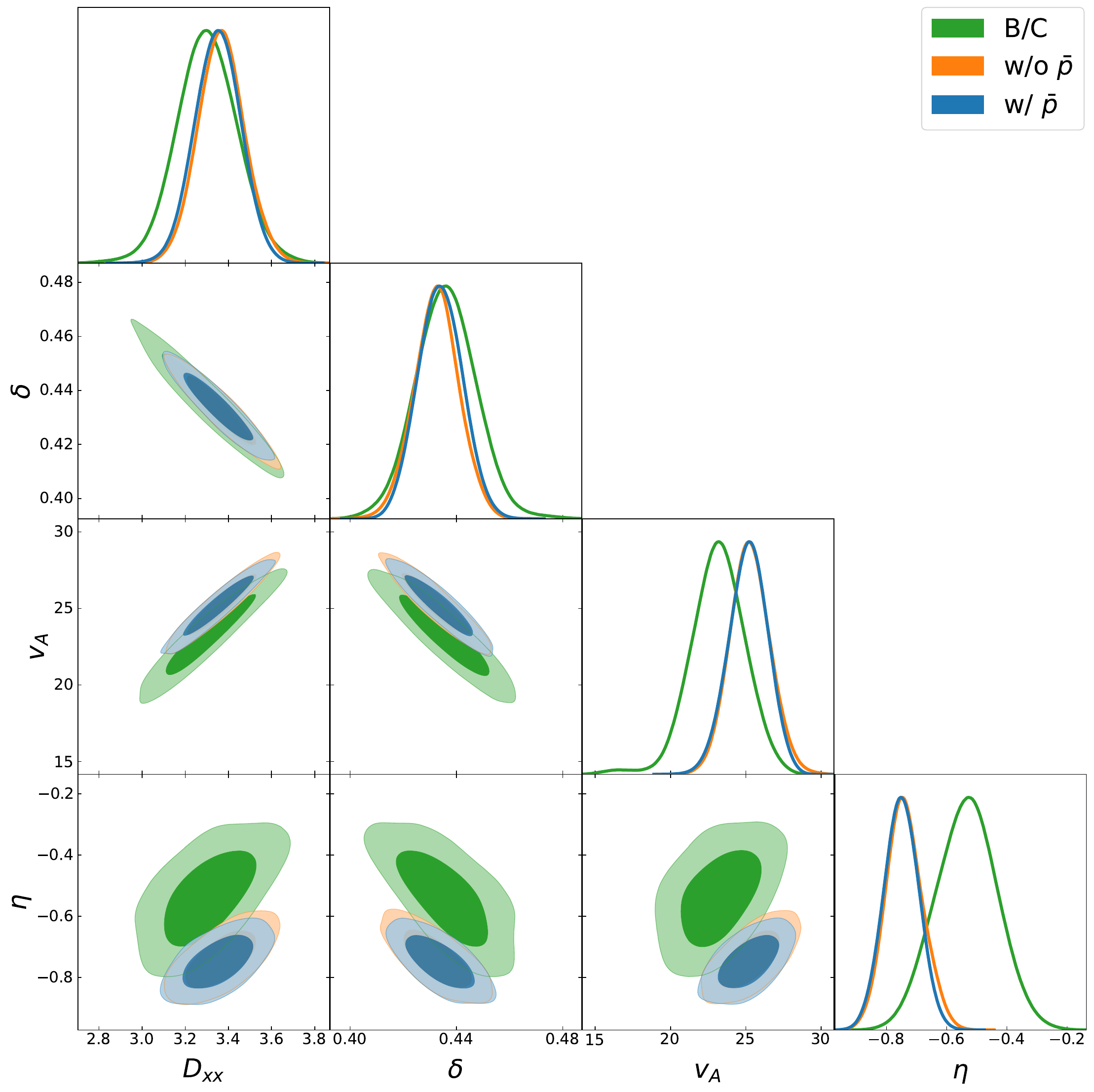}}
    \end{center}
    \captionsetup{justification=raggedright}
    \caption{The 1-D and 2-D posterior distributions of the transport parameters for different elements. The left and right panels show the results for the halo height $L=7.17$ and 3.44 kpc, respectively. The fitting outcomes of step one employing B/C and B/O are represented by the green line. The blue and orange lines represent the results of step two utilizing the proton, Helium, and antiproton data, and the outcomes obtained from the proton and Helium data alone, respectively.}\label{fig:posterior}
\end{figure*}

\subsection{\label{sec:mcmc}Computation of a WIMP contribution}%
Finally, we conduct an MCMC analysis to investigate the potential annihilation of DM particles, which are WIMPs in this case, into hadronic states that generate $\bar{p}$. The purpose of this analysis is to estimate the mass and $\langle\sigma v\rangle$ of DM, that can account for the possible observed inconsistencies when assuming a purely secondary origin of antiprotons. Specifically, we focus on the $\chi\chi\rightarrow b\bar{b}$ annihilation channel, as other quark final state channels yield similar contributions to the production of antiprotons~\cite{Cuoco:2017rxb, caloreAMS02AntiprotonsDark2022}. The corresponding source term is given by
\begin{equation}
q_{\bar{p}}^{\mathrm{DM}}=\frac{\langle\sigma v\rangle}{2 m_\chi^2} \frac{\mathrm{d} N}{\mathrm{~d} E} \rho(\boldsymbol{x})^2\;,
\end{equation}
where the factor 1/2 corresponds to the DM particle being scalar or Majorana fermion,
$m_{\chi}$ is the mass of the DM particle, 
$\langle\sigma v\rangle$ is the
thermally averaged DM annihilation cross section, and $\mathrm{d}N/\mathrm{d}E$ is the antiproton
production spectrum per annihilation, which are given by PPPC 4 DM ID~\cite{Cirelli:2010xx}. The DM density profile $\rho(\boldsymbol{x})$
is taken to be the Navarro-Frenk-White
distribution~\cite{Navarro:1996gj}
\begin{equation}
\rho(r)=\frac{\rho_s}{\left(r / r_s\right)\left(1+r / r_s\right)^2}\;,
\end{equation}
with $r_s = 20$ kpc and $\rho_s = 0.35$ GeV. This choice of parameters corresponds to 
a local DM density of 0.4 GeV $\text{cm}^{-3}$, which is consistent with the recent constraints from the Galactic rotation curve~\cite{Karukes:2019jxv, Benito:2019ngh}. Alternative density profiles, such as the Einasto~\cite{Graham:2005xx,Navarro:2008kc} or Burkert~\cite{Burkert:1995yz, Salucci:2000ps} profiles, are not taken into account in our study, due to their negligible impact on the final antiproton flux, as evidenced in Ref. ~\cite{Cuoco:2016eej, Cuoco:2017iax}.

\section{RESULTS \& DISCUSSION}\label{sec results}
\subsection{Results of fiducial analysis}
Table \ref{bc} displays the outcome of the fitting process in step one. Our calculation for the B/C and B/O ratios fit well with the AMS-02 measurement, as evident by the small $\chi^2$ statistic. It is noteworthy that the optimal scaling factors for the various cross sections associated with secondary production are confined within the $2\sigma$ threshold established by the cross section data, with all factors exhibiting a variation of less than 5\%. As a result, the impact of secondary production cross sections on the propagation parameters is negligible and does not significantly alter the central values of propagation parameters. Their primary effect on the propagation parameters is the slight broadening of their error bars.

\begin{table}[!h]
\captionsetup{justification=raggedright}
\caption{\label{bc} Posterior mean and 95\% credible uncertainties of the model parameters and $\chi^2$ value}
\begin{center}
\begin{tabular} { l  c c}

  Parameter &  L=7.17 kpc & L=3.44kpc\\
\hline
{$D_{xx}(10^{28} \text{cm}^2\text{s}^{-1})         $} & $6.03^{+0.50}_{-0.45}      $& $3.30^{+0.27}_{-0.27}      $\\

{$\delta         $} & $0.440^{+0.021}_{-0.022}   $ & $0.436^{+0.022}_{-0.022}   $\\

{$v_{A}          $(km/s)} & $21^{+3}_{-3}              $& $23^{+4}_{-3}              $\\

{$\eta           $} & $-0.48^{+0.25}_{-0.25}     $& $-0.53^{+0.20}_{-0.20}     $\\
\hline

{$XS_{\ce{^{12}C}\to \ce{^{11}B}}         $} & $0.961^{+0.033}_{-0.034}   $& $0.959^{+0.037}_{-0.036}   $\\

{$XS_{\ce{^{12}C} \to\ce{^{11}C}\to\ce{^{11}B} }          $} & $1.013^{+0.015}_{-0.018}   $& $1.013^{+0.015}_{-0.015}   $\\

{$XS_{\ce{^{12}C}\to \ce{^{10}B}}          $} & $1.023^{+0.063}_{-0.065}   $& $1.019^{+0.061}_{-0.058}   $\\

{$XS_{\ce{^{16}O}\to \ce{^{10}B}(\ce{^{10}C})}           $} & $1.041^{+0.063}_{-0.063}   $& $1.044^{+0.059}_{-0.058}   $\\

{$XS_{\ce{^{16}O}\to \ce{^{11}B}(\ce{^{11}C})}           $} & $0.973^{+0.053}_{-0.055}   $& $0.979^{+0.050}_{-0.050}   $\\

{$XS_{\ce{^{14}N}\to \ce{^{11}B}(\ce{^{11}C})}           $} & $1.026^{+0.057}_{-0.053}   $& $1.028^{+0.056}_{-0.054}   $\\

{$XS_{\ce{^{15}N}\to \ce{^{11}B}}          $} & $1.009^{+0.090}_{-0.086}   $& $1.008^{+0.088}_{-0.093}   $\\
\hline

$\chi^2/\text{d.o.f.}                    $ & $96.56/123   $ & $98.3/123   $\\

\hline
\end{tabular}

\end{center}
\end{table}

Fig. \ref{fig:posterior} is the triangle plot of the fitting results for step two and shows the 1D marginalized posterior probability density
functions of the parameters and 2D contour plots of 68\%
and 95\% credible regions for all the combinations. 
The injection parameters are omitted for simplicity.
It is evident that most of the propagation parameters are consistent with those obtained during step one, except for $\eta$. Notably, the outcomes attained in the presence or absence of $\bar{p}$ are almost indistinguishable, differing only in a negligible discrepancy in the width of their respective posterior distributions. This similarity suggests that the impact of $\bar{p}$ on the overall results is minimal, indicating the internal consistency of the study. 

The first two columns of Table \ref{table: pbar} present a summary of the results, listing the mean values and posterior 95\% ranges. The propagation parameters with and without antiproton exhibit full consistency within uncertainties for both $L$ values. The halo diffusion slope indexes $\delta$ fall within the range of the Kolmogorov type with $\delta=1/3$ and the Kraichnan type with $\delta=1/2$. The fitted scale factors $c_{\bar{p}}$ are within the experiment's uncertainty of $\sigma_{\mathrm{norm.}}\simeq0.08$ as given in Ref.~\cite{Korsmeier:2018gcy}. Regarding solar modulation, $\phi^{-}_{0}$ is smaller than $\phi^{-}_{0}$ by $\mathcal{O}(0.1)$ GV. Confirmation of this result must wait for the monthly antiproton flux's release by the AMS-02 collaboration.

It is worth mentioning that the value of $\chi^2_{\mathrm{min}}/\mathrm{DoF}$ is significantly lower than 1. This may be due to that we employ a quadratic combination of the AMS-02 statistical and systematic errors, thereby neglecting the correlations present within the AMS-02 systematic errors. Nevertheless, it has been previously suggested that the inclusion of these correlations, as proposed in Refs. ~\cite{Boudaud:2019efq, Heisig:2020nse}, would result in even flatter residues, which does not alter the conclusions drawn in this study. However, future investigations that account for error correlations are contingent upon the AMS-02 collaboration publishing their covariance matrix~\cite{Derome:2019jfs}.

\begin{figure}[htbp]
\includegraphics[width=0.48\textwidth]{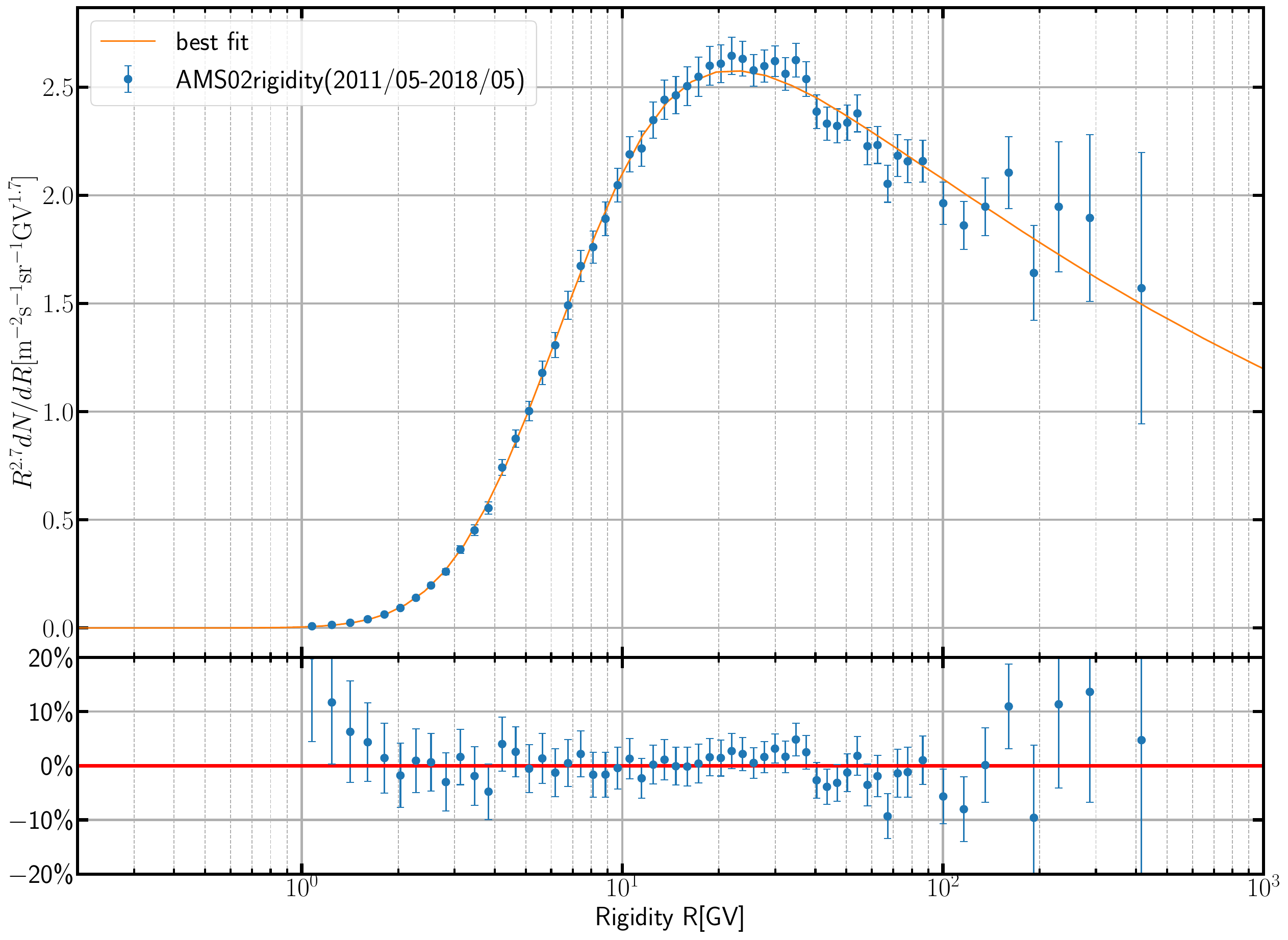}\\
\includegraphics[width=0.48\textwidth]{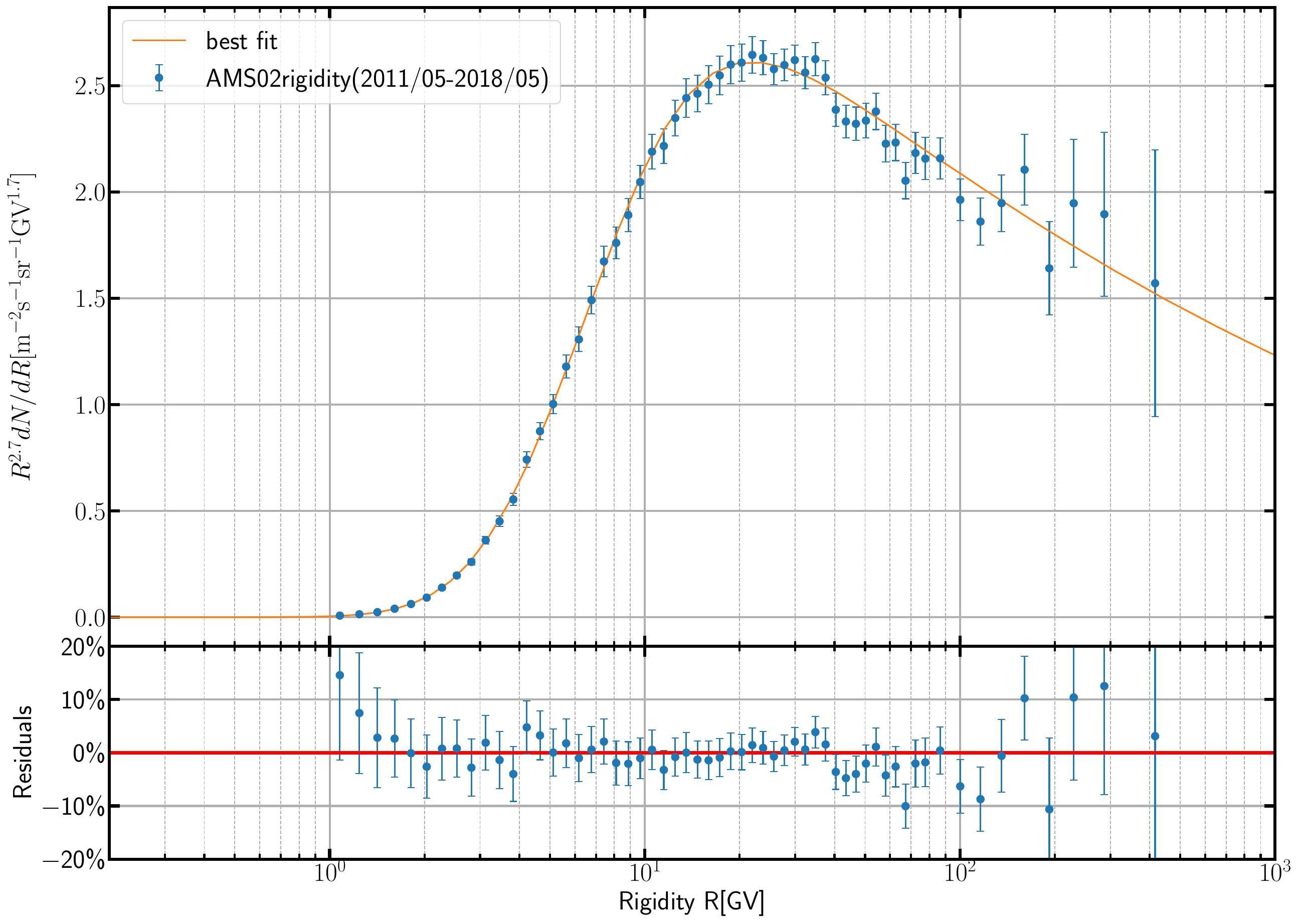}\\
\captionsetup{justification=raggedright}
\caption{Best-fit model predictions of the antiproton fluxes, compared with the AMS-02 data~\cite{AMS:2021nhj}. The top and bottom panels show the results for the thick halo with $L=7.17$ kpc and thin halo with $L=3.44$ kpc, respectively.\label{fig:pbar}}
\end{figure}

The $\bar{p}$ spectrum, calculated using the propagation parameters determined in step two, is depicted in Fig. \ref{fig:pbar}. The spectrum is in excellent agreement with the corresponding AMS-02 data in most energy ranges. However, a slight discrepancy is observed at the low energy end of the spectrum, which may be attributed to the over-simplified solar modulation model. Furthermore, a few data points at the high energy end exceed our calculation. Nevertheless, the large error bars associated with these data points prevent drawing any meaningful conclusions.

The results obtained above lead to two significant conclusions. 
Firstly, 
we think that the exceptional precision of the proton and Helium measurements appears to offer a partial resolution to the degeneracy~\cite{Jin:2014ica} traditionally observed between source and propagation parameters. A full degeneracy for these parameters in proton and helium would imply a significant overlap between the posterior distribution of propagation parameters and the prior distribution derived from B/C. Instead, we can see from Fig~\ref{fig:posterior} that they are in fact misaligned, and the effect of adding secondary element, in this case antiproton, has littele effect on the posterior distribution. This is an intersting finding as it suggests that propagation parameters can be effectively fixed using primary particles alone, thereby mitigating the uncertainties associated with secondary production cross sections.
Secondly, the antiproton spectrum observed from AMS-02 aligns entirely with a secondary astrophysical origin, suggesting that no further primary antiproton sources are required to explain the AMS-02 data. 

\subsection{Impact of hadronic interaction models}
\begin{figure}[htbp]
\includegraphics[width=0.48\textwidth]{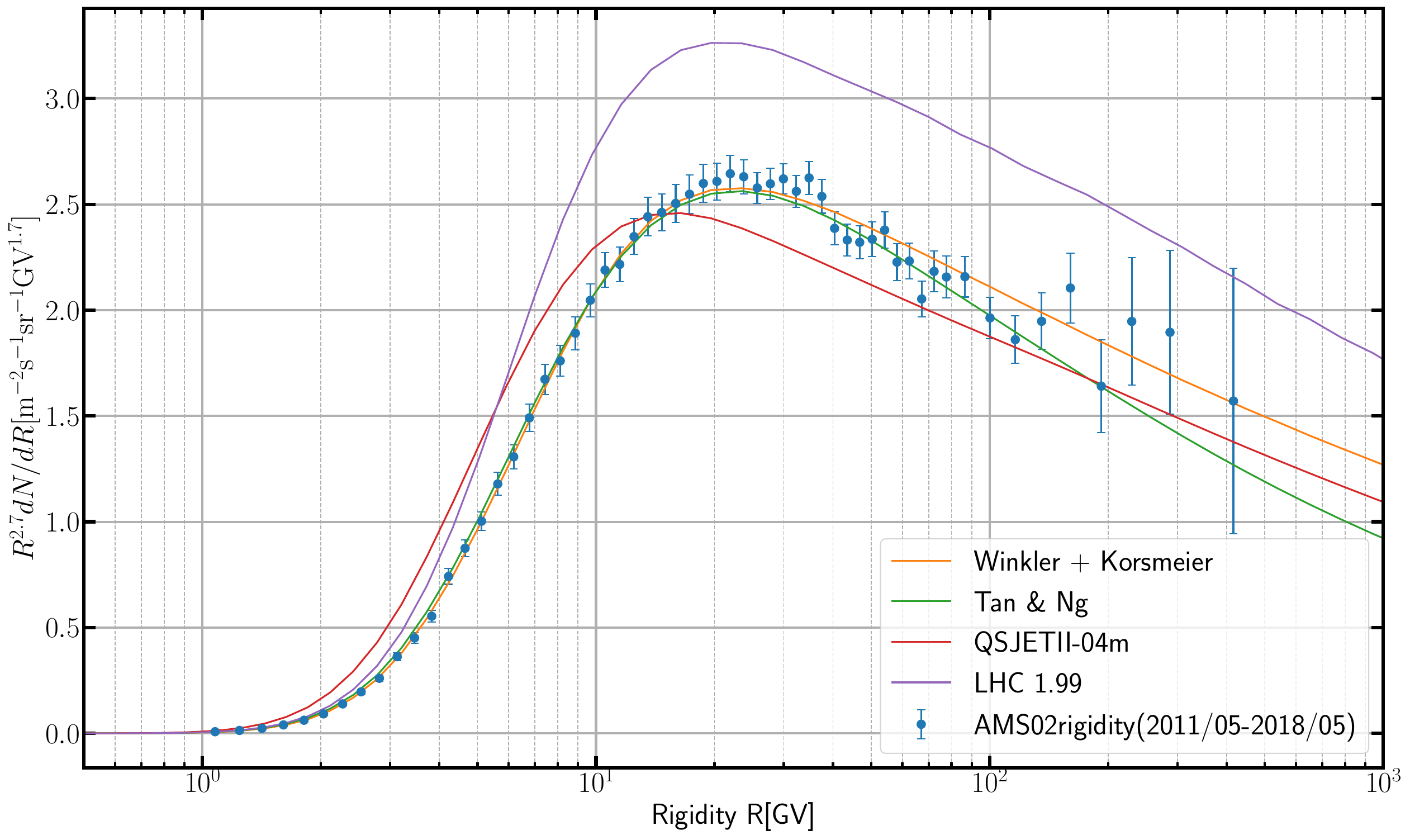}\\
\captionsetup{justification=raggedright}
\caption{The $\bar{p}$ flux expected from different hadronic models for $L=7.17$ kpc, in comparison with the AMS-02 measurement~\cite{AMS:2021nhj}. Four lines represent the results for the models of Winkler + Korsmeier (orange), LHC 1.99 (purple), QSJET II-04 m (red), and Tan \& Ng (green), respectively.
The parameters used here give the best fit to the proton and Helium data. The solar modulation potentials are set to be $\phi^{-}_0=0.56$GV,
$\phi^{-}_1=0.40$GV.\label{fig:pbar_hadronic}}
\end{figure}

In this subsection, various hadronic models are tested to determine their efficacy in antiproton production cross section. Three other hadronic models, namely LHC 1.99~\cite{Pierog:2009zt}, QSJET II-04 m~\cite{Kachelriess:2015wpa}, and Tan \& Ng~\cite{Tan:1983kgh}, are considered for $L=7.17$ kpc here. We firstly plot the \textit{predictions} of different hadronic interaction models in Fig. \ref{fig:pbar_hadronic}, utilizing the parameters established from the fit to the protons and Helium data. It is clear from the figure that the two phenomenological models, LHC 1.99 and QSJET II-04m, are unable to accurately replicate the $\bar{p}$ data. In contrast, the two empirical models, Winkler + Korsmeier and Tan \& Ng, produce comparable results that align with the data within experimental error in most energy ranges. It should be noted that the GALPROP default Tan \& Ng model appears to underestimate the $\bar{p}$ flux at the high energy end, with notable discrepancies compared to the last several data points.

\begin{figure}[htbp]
\includegraphics[width=0.48\textwidth]{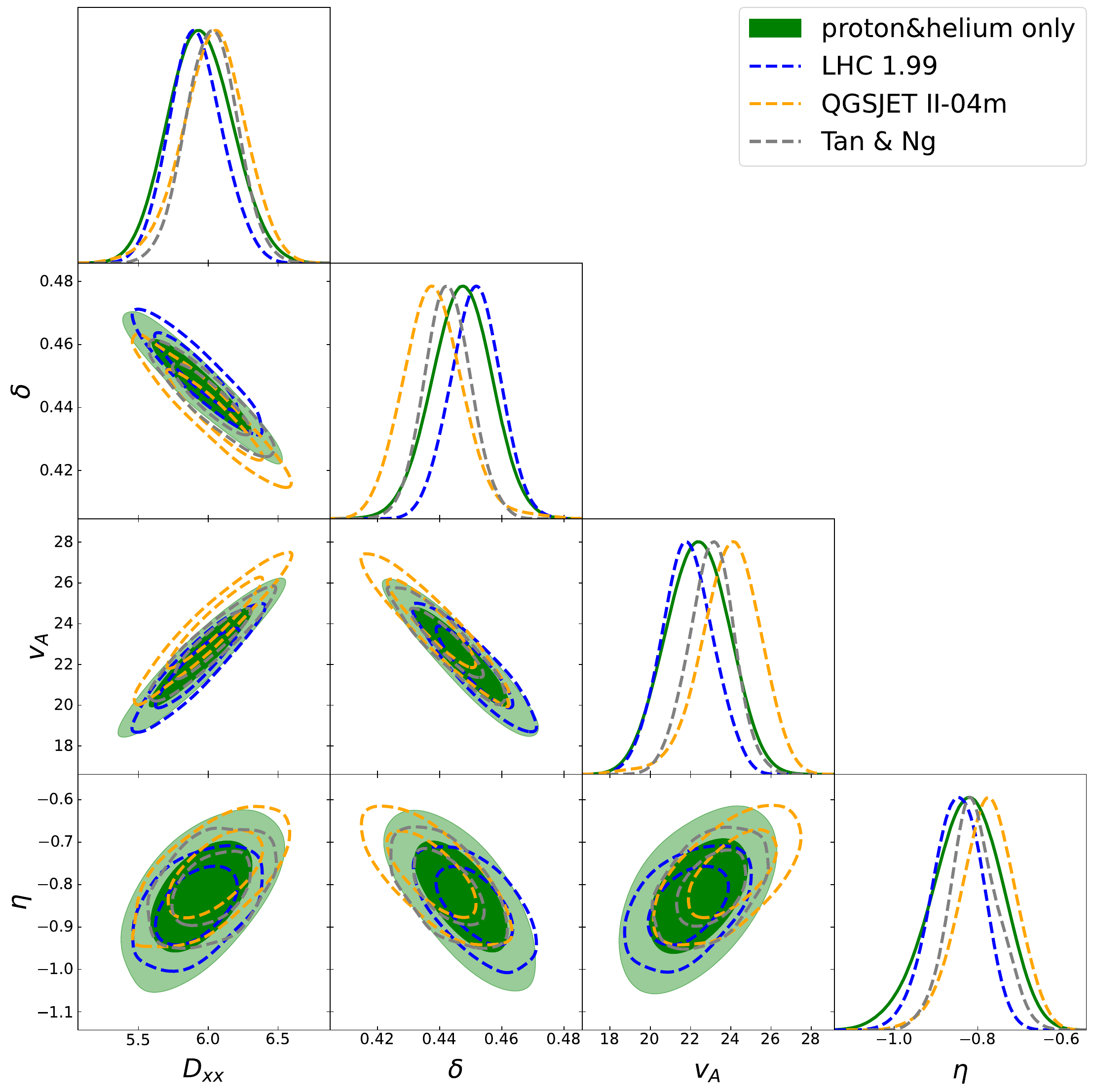}\\
\captionsetup{justification=raggedright}
\caption{Posterior distribution of propagation parameters using different hadronic models for halo height $L=7.17$ kpc. The solid line represents the fitting outcomes obtained exclusively from the proton and Helium data, while the dashed lines depict the fitting results with antiproton data, which are calculated using the models proposed by LHC 1.99 (blue), QSJET II-04 m (yellow), and Tan \& Ng (gray).\label{fig:posterior_hadronic} }
\end{figure}

Subsequently, various MCMC fits are performed utilizing these models. The results are presented in Figure \ref{fig:posterior_hadronic}. Our examination of Figure \ref{fig:pbar_hadronic} is reinforced by the analysis, which demonstrates that the phenomenological models are less effective in comparison to the empirical ones. Notably, the propagation parameters with and without antiproton are no longer entirely consistent, with the greatest discrepancy occurring when the QSJET II-04m model is employed. The results of the fits are presented in the final column of Table \ref{table: pbar}. It is evident that the $\chi^2$ statistic significantly worsens when compared to that of the empirical models. Furthermore, the two phenomenological models display a scale factor of $c_{\bar{p}}$ approximately 20\%, which is considerably higher than the 5\% scale factor observed in the empirical models. Moreover, it is worth noting that the QSJET II-04m model requires a solar modulation potential greater than 1GV, which could be unphysical.

Concerning the comparison between the two empirical models, the latest Winkler + Korsmeier model that incorporates the collider data is slightly favored over the Tan \& Ng model. This can be seen from the fact that the propagation parameters derived with the Tan \& Ng model demonstrate a slight discrepancy from the fit involving only the proton and Helium data, necessitating a larger $c_{\bar{p}}$ of 5\% in contrast to 2\%. Nevertheless, these divergences are all contained within the 2$\sigma$ uncertainty limits, thus we cannot assert definitively that Winkler + Korsmeier is the preferred model.

\begin{table*}[t]
  \centering
    \captionsetup{justification=raggedright}
  \caption{The mean values and posterior 95\% ranges of key parameters in different scenarios. For the results of the alternative hadronic models listed in the last three columns, $L$ is taken to be 7.1.7 kpc.}
\begin{tabular}{cccccccccc}
\hline
                                & \multicolumn{2}{c}{Standard Analysis}                       &  & \multicolumn{2}{c}{w/o $\bar{p}$}                           &  & \multicolumn{3}{c}{Hadronic Models}                                                        \\ \cline{2-3} \cline{5-6} \cline{8-10} 
                                & $L=7.17$ kpc                 & $L=3.44$ kpc                 &  & $L=7.17$ kpc                 & $L=3.44$ kpc                 &  & Tan \& Ng                    & EPOS 1.99                    & QGSJETII-04m                 \\ \hline
$c_{\bar{p}}$                   & $1.019^{+0.036}_{-0.034}   $ & $1.002^{+0.033}_{-0.032}   $ &  & -                            & -                            &  & $1.050^{+0.034}_{-0.033}   $ & $0.824^{+0.030}_{-0.028}   $ & $1.209^{+0.040}_{-0.039}   $ \\
$\phi^{+}_{0}$(GV) & $0.672^{+0.057}_{-0.053}   $ & $0.735^{+0.054}_{-0.058}   $ &  & $0.667^{+0.061}_{-0.065}   $ & $0.736^{+0.058}_{-0.053}   $ &  & $0.677^{+0.047}_{-0.047}   $ & $0.652^{+0.053}_{-0.053}   $ & $0.708^{+0.064}_{-0.064}  $ \\
$\phi^{-}_{0}$(GV) & $0.54^{+0.10}_{-0.11}      $ & $0.49^{+0.10}_{-0.11}      $ &  & -                            & -                            &  & $0.627^{+0.097}_{-0.10}    $ & $0.654^{+0.094}_{-0.10}    $ & $1.20^{+0.10}_{-0.10}$       \\
$\phi^{-}_{1}$(GV) & $0.27^{+0.25}_{-0.27}      $ & $< 0.348              $      &  & -                            & -                            &  & $0.43^{+0.29}_{-0.27}      $ & $< 0.203              $      & $0.98^{+0.32}_{-0.30}$         \\ \hline 
$\chi^2$/DoF                        & 73.6/183                         &      65.1/183                        &  &             49.2/128                 &                         43.1/128     &  &              70.0/183                &                93.4/183              &         99.3/183                     \\ \hline 
$D_{xx}(10^{28} \text{cm}^2\text{s}^{-1})$                           & $5.96^{+0.42}_{-0.38}      $ & $3.35^{+0.21}_{-0.21}      $ &  & $5.94^{+0.47}_{-0.44}      $ & $3.37^{+0.22}_{-0.21}      $ &  & $6.02^{+0.36}_{-0.36}      $ & $5.91^{+0.38}_{-0.36}      $ & $6.03^{+0.43}_{-0.48}      $ \\
$\delta$                        & $0.446^{+0.016}_{-0.017}   $ & $0.434^{+0.016}_{-0.016}   $ &  & $0.447^{+0.018}_{-0.019}   $ & $0.432^{+0.017}_{-0.017}   $ &  & $0.443^{+0.015}_{-0.014}   $ & $0.451^{+0.015}_{-0.016}   $ & $0.438^{+0.019}_{-0.019}   $ \\
$v_{A}$(km/s)                           & $23^{+3}_{-2}              $ & $25.2^{+2.5}_{-2.6}        $ &  & $22^{+3}_{-3}              $ & $25.3^{+2.7}_{-2.5}        $ &  & $23.0^{+2.2}_{-2.3}        $ & $22^{+3}_{-3}              $ & $24^{+3}_{-3}              $ \\
$\eta$                          & $-0.82^{+0.12}_{-0.12}     $ & $-0.75^{+0.11}_{-0.11}     $ &  & $-0.83^{+0.16}_{-0.17}     $ & $-0.74^{+0.13}_{-0.12}     $ &  & $-0.81^{+0.12}_{-0.12}     $ & $-0.85^{+0.11}_{-0.13}     $ & $-0.78^{+0.13}_{-0.14}     $ \\ \hline  
\end{tabular}
  \label{table: pbar}
\end{table*}

In conclusion, our findings show that the two empirical models are strongly preferred over the two phenomenological models, which are already superior to other phenomenological models in  reproducing collider data in the low energy region.
This finding also highlights the potential for CR data to be utilized as a means for selecting the optimal hadronic models. Furthermore, with a deeper understanding of the propagation mechanism and access to more accurate CR data, the possibility of determining nuclear production cross section via CRs is becoming increasingly plausible.

\subsection{Testing possible antiproton production from DM}

The AMS-02 antiproton data exhibits a high degree of agreement with our calculations, thereby resulting in a robust constraint on DM annihilation. We utilize the standard Bayes analysis protocol to ascertain the corresponding 95\% upper limit for a specific value of $m_{\chi}$ through the following equation~\cite{brooksHandbookMarkovChain2011}
\begin{equation}
\frac{\int_0^{\langle\sigma v\rangle_{95}} \mathcal{P}(x) \mathrm{d} x}{\int_0^{\infty} \mathcal{P}(x) \mathrm{d} x}=0.95\;.
\end{equation}

\begin{figure}[htbp]
\includegraphics[width=0.46\textwidth]{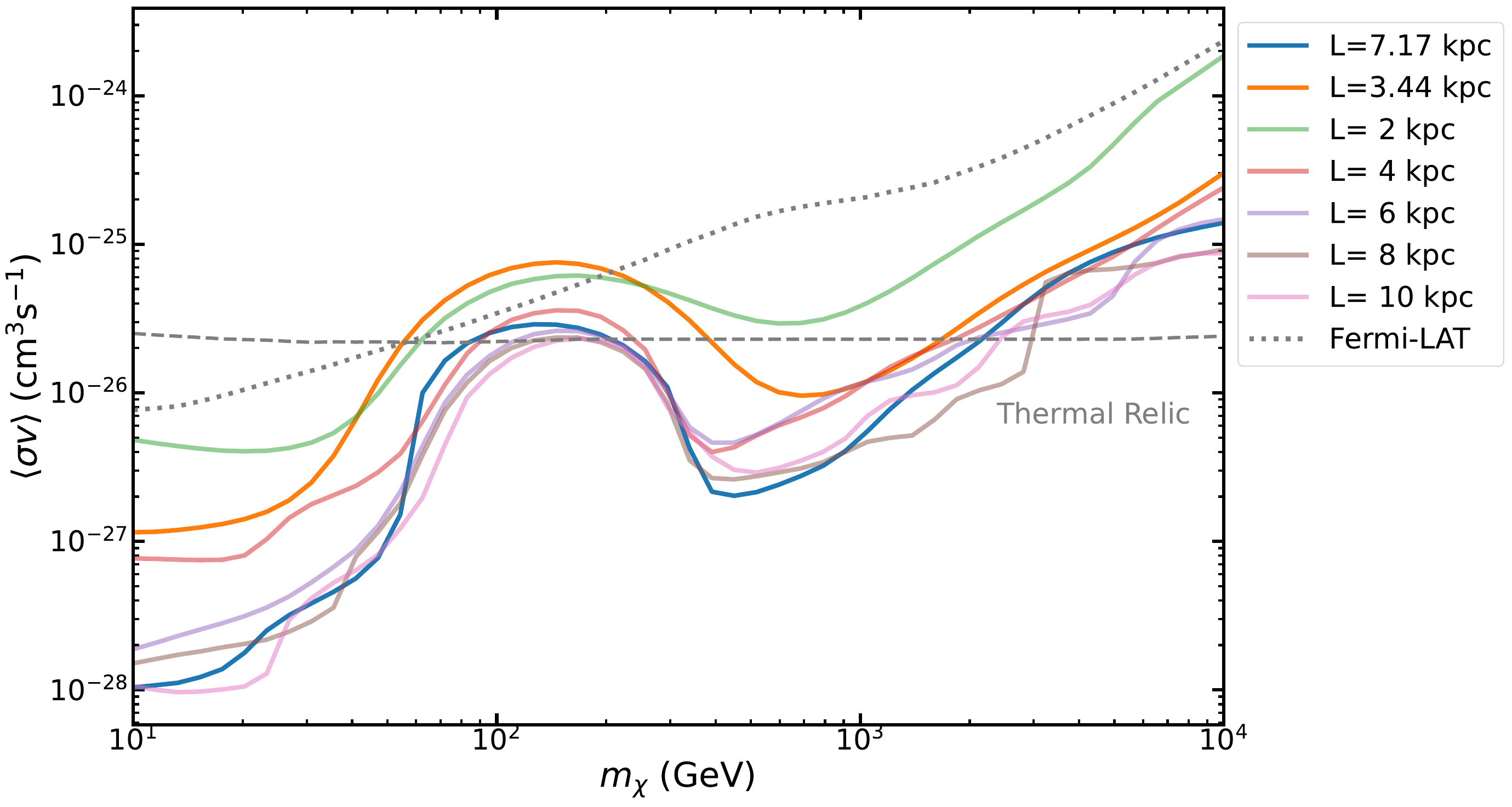}\\
\captionsetup{justification=raggedright}
\caption{
The 95\% upper limits on the DM annihilation cross section $\langle\sigma v\rangle$ derived from the AMS-02 $\bar{p}$ data for different choice of halo sizes. For comparison, we also show the upper limits from the Fermi-LAT observation of dwarf galaxies~\cite{Fermi-LAT:2016uux} as the grey dotted line. The dashed gray curve corresponds to the thermal relic cross section from Ref. ~\cite{Steigman:2012nb}.}\label{fig:pbar_lim}
\end{figure}

The results are shown in Fig. \ref{fig:pbar_lim}. for different choices of halo height.
The limits derived from the investigation are demonstrated to be typically stronger in comparison to the outcomes of gamma-ray observations of dwarf galaxies~\cite{Fermi-LAT:2016uux}. However, it should be noted that the effect of varying halo sizes is not a mere shift of the exclusion limit as naively expected. Instead, the line shape of the limit undergoes moderate changes, 
particularly in the lower energy spectrum, where the effect of reacceleration, the $\eta$ parameter, and solar modulation are most pronounced.
Nonetheless, the qualitative conclusion that the DM annihilation cross section increases as the halo thickness decreases remains valid. Additionally, it is important to mention that the constraints established in this study may be subject to a constant factor reduction (increase) if the local density of DM is proven to be higher (lower)~\cite{deSalas:2020hbh}.

\begin{table}[h]
\begin{center}
\captionsetup{justification=raggedright}
\caption{\label{table dm}DM parameters, $\chi^2$ values, number of free fit parameters,  and significance for the best fits. The total $\chi^2$ values refer to the fits with (without) DM.}
\begin{tabular} { l  c c}

  Parameter &  L=7.17 kpc & L=3.44kpc\\
\hline
{$m_{\chi}$(GeV)} & $132$& $166$\\

{$\langle\sigma v\rangle(10^{-26}\text{cm}^3\text{s}^{-1}$)} & $0.98$ & $1.71$\\
\hline
{$\chi^2_{\text{tot}}$(200 points)} & $65.1 (73.6)$& $59.9 (65.1)$\\

{No. of fit param.} & $19 (17)$& $19 (17)$\\

{$\Delta\chi^2$} & $8.5$& $5.2$\\

{local sig.} & $2.7\sigma$& $2.1\sigma$\\

\hline
\end{tabular}
\end{center}
\end{table}

Table \ref{table dm} presents the properties of DM as inferred from fitting, with the local significance of the signal determined using the likelihood ratio test. It should be noted that the global significance of the DM signal is lower than the provided local significance due to the impact of the look-elsewhere effect (LEE)~\cite{Gross:2010qma}. Because the local evidence for a DM signal is statistically insignificant, we refrain from conducting a computationally demanding evaluation of its global significance. Nonetheless, according to Ref. ~\cite{Heisig:2020nse}, the global significance of the signal was estimated to be approximately $0.8\sim1\sigma$ lower than its local significance.

\subsection{Impact of new AMS-02 data\label{subsec:new data}}
\begin{figure}[htbp]
\includegraphics[width=0.5\textwidth]{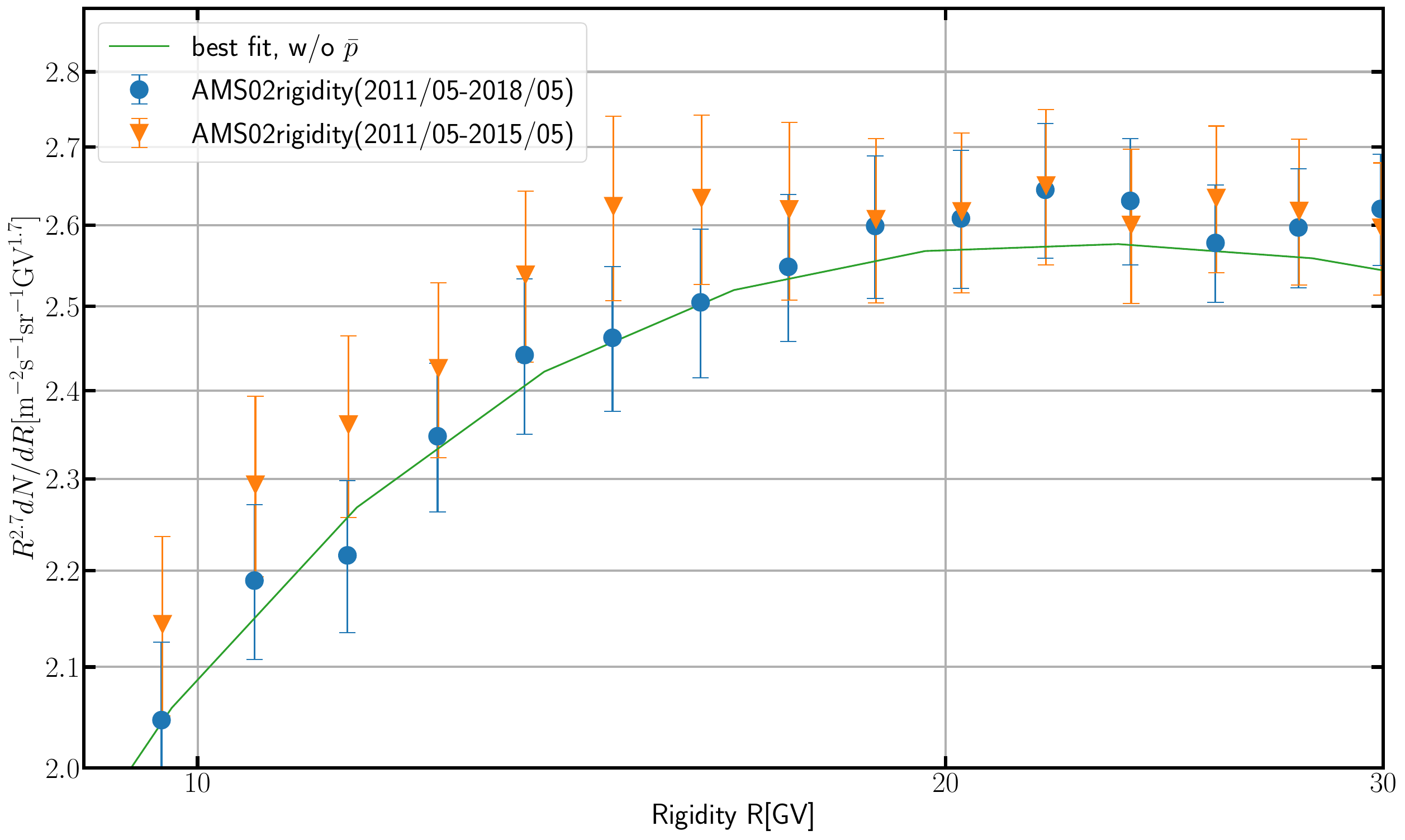}\\
\captionsetup{justification=raggedright}
\caption{The subtle difference between the AMS-02 data taken during 2011-2015~\cite{AMS:2016oqu} and 2011-2018~\cite{AMS:2021nhj}\label{fig:oldData} in the energy range of 10-20 GV. The solid green line is the best-fit spectrum to the data taken during 2011-2018.}
\end{figure}

The derived DM mass accounting for the potential antiproton excess, which falls within the range of 130-160 GeV, is significantly greater than that observed in prior investigations~\cite{Cui:2018klo, Cuoco:2019kuu, Heisig:2020jvs} and is incompatible with the gamma-ray Galactic Center excess ~\cite{DiMauro:2021qcf}. The study attributes this result to the use of the most recent AMS-02 antiproton data acquired between 2011 and 2018~\cite{AMS:2021nhj}. Fig. \ref{fig:oldData} illustrates the comparison of the old and new antiproton data sets, demonstrating that while both data sets exhibit consistency within their respective margins of error, there exist significant discrepancies in the central values of data points between 10 and 20 GV. Specifically, the bump structure evident in the old data, which has previously been a subject of controversy in the context of the antiproton excess, is absent in the new data.

\begin{figure}[htbp]
\includegraphics[width=0.48\textwidth]{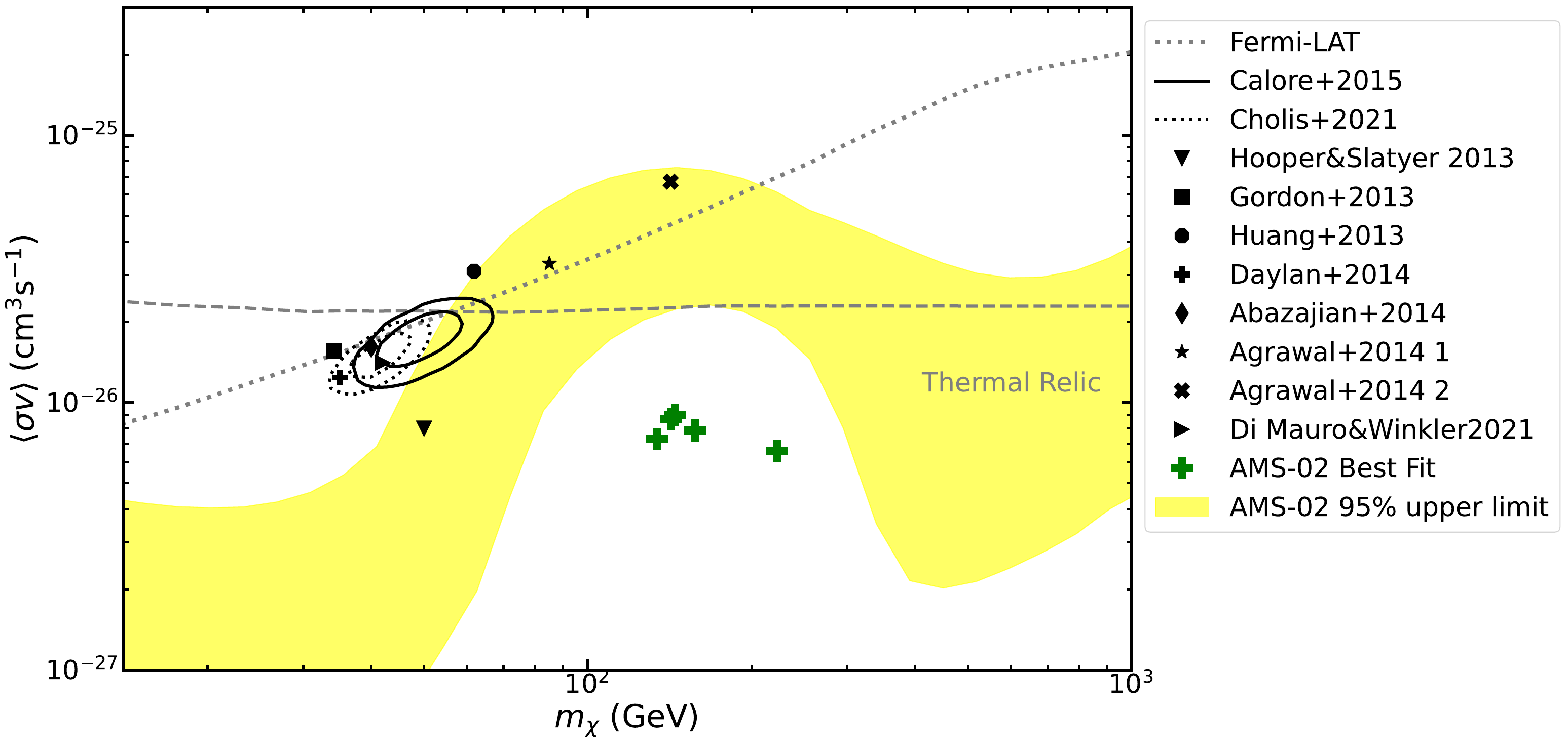}\\
\includegraphics[width=0.48\textwidth]{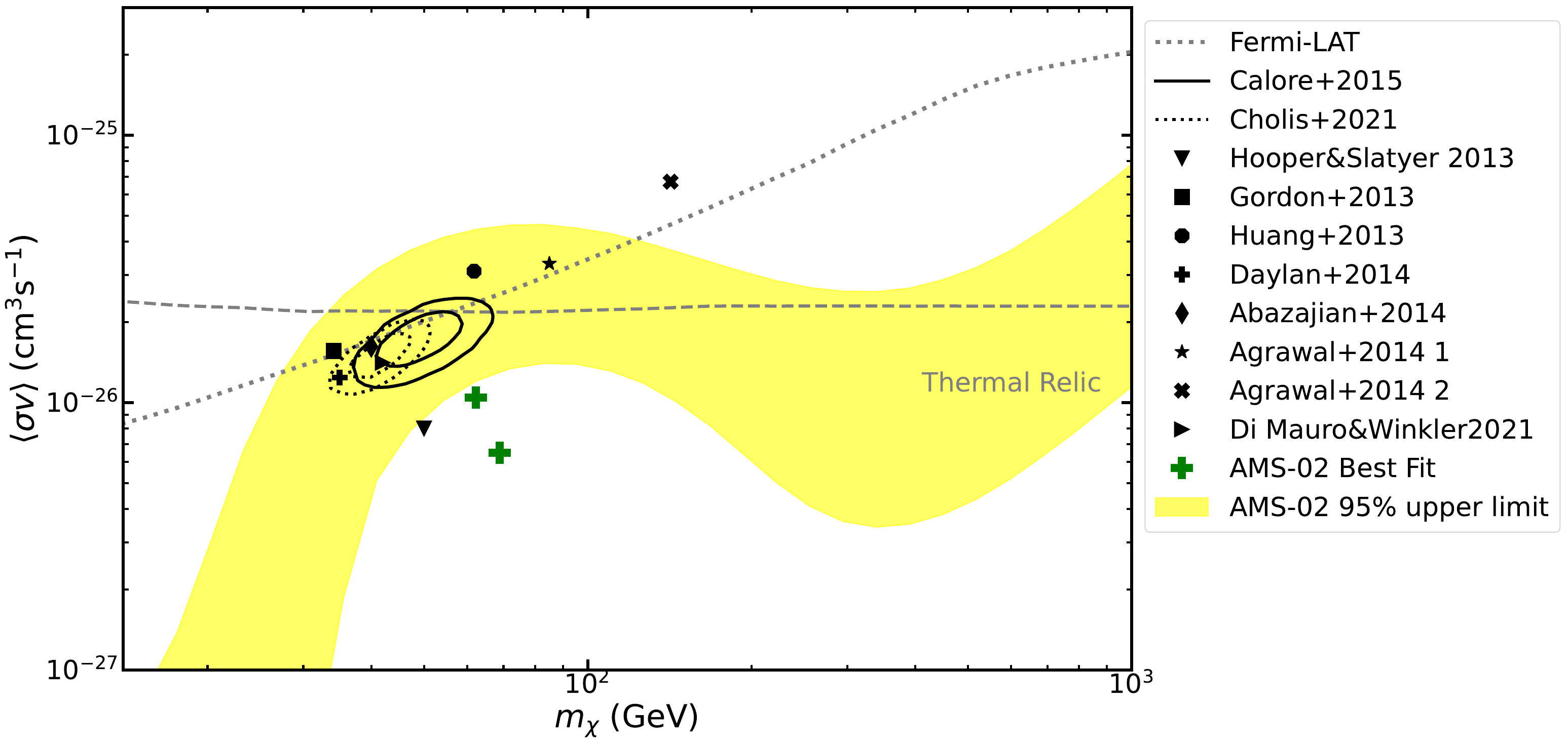}\\
\captionsetup{justification=raggedright}
\caption{
The 95\% upper limits on the DM annihilation cross section derived from AMS-02 antiproton data, illustrated as a yellow band, alongside the best-fit values obtained from various GCE analyses, marked by black points. The upper panel corresponds to analyses conducted with the most recent AMS-02 data~\cite{AMS:2021nhj}, while the lower panel presents results derived from earlier AMS-02 data~\cite{AMS:2016oqu}. Additionally, confidence intervals from two representative GCE studies are included, highlighting the typical observed range in GCE research. Green crosses indicate the best-fit values from antiproton analysis for different halo sizes. For comparison, the upper limit from the Fermi-LAT observations of dwarf galaxies is depicted as the dotted line~\cite{Fermi-LAT:2016uux}. The thermal relic cross section, as reported in Ref.~\cite{Steigman:2012nb}, is shown as the dashed line.}\label{fig:dm_oldData}
\end{figure}

{
We further perform MCMC fits using the old data, focusing on two extreme halo sizes: 2 kpc and 10 kpc.} As shown in Fig. \ref{fig:dm_oldData}, the results of the MCMC analysis align with the observations described above. Specifically, the estimated mass of the hypothetical DM signal is in agreement with the findings of previous studies conducted using the old data.

The results of the current investigation disfavor the feasibility of utilizing a signal DM model to explain both the GCE and the antiproton excess. {
To demonstrate this, in Fig.~\ref{fig:dm_oldData} we plot various best-fit DM parameters and representative confidence intervals that can explain the GCE in the literature\cite{Hooper:2013rwa,Gordon:2013vta,2013arXiv1307.6862H,Daylan:2014rsa,Abazajian:2014fta,Agrawal:2014oha,Calore:2014xka,Cholis:2021rpp,DiMauro:2021qcf} alongside our best-fit values from antiproton analyses. Notably, DM masses derived using the latest AMS-02 data predominantly exceed those required to explain the GCE, with the exception of findings reported in Ref.~\cite{Agrawal:2014oha}. Additionally, many DM parameter regions explaining the GCE could be effectively ruled out at the 95\% confidence level, even the large uncertainties of the halo size are condiered. Conversely, the best-fit DM masses from the older AMS-02 data align with those accounting for the GCE, in agreement with eariler studies~\cite{Cui:2016ppb, Cuoco:2017rxb}.} This limitation also pertains to the efforts to address the W-boson mass anomaly while simultaneously accounting for the antiproton excess in Ref ~\cite{Zhu:2022tpr}. This discovery is in qualitative accordance with the findings presented in Ref. ~\cite{Luque:2021ddh}.

\section{SUMMARY\label{sec:conclusion}}
The current research endeavor involves the evaluation of the CR antiproton spectrum, which is then compared with the most recent AMS-02 results. Given the absence of any discernible excess in the $\bar{p}$ spectrum, the study subsequently establishes upper limits on the annihilation cross section of DM.

In this study, the DR2 propagation model is selected due to its preference for the latest secondary-to-primary ratios from the AMS-02 experiment. Using the latest AMS-02 B/C and B/O data and incorporating uncertainties in the Boron production cross section, the constraints on the propagation parameters are updated. The injection spectrum for proton and Helium is then determined via a combined fit to the AMS-02 proton, Helium, and antiproton data. The fitting process incorporates a charge-sign dependent solar modulation model and the most current $\bar{p}$ production cross section model. Alternative $\bar{p}$ production models that do not incorporate the most recent collider data are also considered, and the effect of using the latest AMS-02 $\bar{p}$ data is evaluated.

The most important findings of this work are summarized as follows:
\begin{itemize}
  \item The $\bar{p}$ spectrum is consistent with a secondary origin, and therefore no contribution from DM is necessary.
  \item Accounting for the various systematic uncertainties, the derived upper bound on the DM annihilation cross section, calculated using the $\bar{p}$ data, could be more stringent than the corresponding constraint obtained from the Fermi observation of the dwarf galaxies.
  \item The utilization of the most recent AMS-02 $\bar{p}$ data has a notable impact. Specifically, the plausibility of a self-consistent explanation for the GCE and the antiproton excess is no longer tenable upon the utilization of the recent $\bar{p}$ data.
  \item The precision of the AMS-02 CR data is sufficient to enable the identification of the most appropriate cross section models: the empirical hadronic interaction models are superior to the phenomenological ones.
\end{itemize}

In the future, the modeling of the transport of CRs in the Galaxy can be improved through the incorporation of more precise models, such as the spatial-dependent diffusion~\cite{Zhao:2021yzf}. Utilizing various numerical tools to model solar modulation~\cite{Boschini:2017gic, Kappl:2015hxv} can also improve our understanding of its impact on the $\bar{p}$ flux. Ultimately, these models will facilitate a more comprehensive understanding of the mechanisms driving the $\bar{p}$ flux, which will enable a more accurate understanding of the nature and properties of DM.

\acknowledgments
This work is supported by the National Natural Science Foundation of China under
the grants No. 12175248, No. 12105292 and No. 2022YFA1604802.

\bibliography{apssamp}

\providecommand{\noopsort}[1]{}\providecommand{\singleletter}[1]{#1}%
\begin{thebibliography}{92}%
\makeatletter
\providecommand \@ifxundefined [1]{%
 \@ifx{#1\undefined}
}%
\providecommand \@ifnum [1]{%
 \ifnum #1\expandafter \@firstoftwo
 \else \expandafter \@secondoftwo
 \fi
}%
\providecommand \@ifx [1]{%
 \ifx #1\expandafter \@firstoftwo
 \else \expandafter \@secondoftwo
 \fi
}%
\providecommand \natexlab [1]{#1}%
\providecommand \enquote  [1]{``#1''}%
\providecommand \bibnamefont  [1]{#1}%
\providecommand \bibfnamefont [1]{#1}%
\providecommand \citenamefont [1]{#1}%
\providecommand \href@noop [0]{\@secondoftwo}%
\providecommand \href [0]{\begingroup \@sanitize@url \@href}%
\providecommand \@href[1]{\@@startlink{#1}\@@href}%
\providecommand \@@href[1]{\endgroup#1\@@endlink}%
\providecommand \@sanitize@url [0]{\catcode `\\12\catcode `\$12\catcode
  `\&12\catcode `\#12\catcode `\^12\catcode `\_12\catcode `\%12\relax}%
\providecommand \@@startlink[1]{}%
\providecommand \@@endlink[0]{}%
\providecommand \url  [0]{\begingroup\@sanitize@url \@url }%
\providecommand \@url [1]{\endgroup\@href {#1}{\urlprefix }}%
\providecommand \urlprefix  [0]{URL }%
\providecommand \Eprint [0]{\href }%
\providecommand \doibase [0]{https://doi.org/}%
\providecommand \selectlanguage [0]{\@gobble}%
\providecommand \bibinfo  [0]{\@secondoftwo}%
\providecommand \bibfield  [0]{\@secondoftwo}%
\providecommand \translation [1]{[#1]}%
\providecommand \BibitemOpen [0]{}%
\providecommand \bibitemStop [0]{}%
\providecommand \bibitemNoStop [0]{.\EOS\space}%
\providecommand \EOS [0]{\spacefactor3000\relax}%
\providecommand \BibitemShut  [1]{\csname bibitem#1\endcsname}%
\let\auto@bib@innerbib\@empty
\bibitem [{AMS()}]{AMS02AlphaMagnetic}%
  \BibitemOpen
  \href {https://ams02.space/} {\bibinfo {title} {{{AMS-02}} | {{The Alpha
  Magnetic Spectrometer Experiment}}}},\ \bibinfo {howpublished}
  {https://ams02.space/}\BibitemShut {NoStop}%
\bibitem [{\citenamefont {Blum}\ \emph {et~al.}(2017)\citenamefont {Blum},
  \citenamefont {Sato},\ and\ \citenamefont {Waxman}}]{Blum:2017iwq}%
  \BibitemOpen
  \bibfield  {author} {\bibinfo {author} {\bibfnamefont {K.}~\bibnamefont
  {Blum}}, \bibinfo {author} {\bibfnamefont {R.}~\bibnamefont {Sato}},\ and\
  \bibinfo {author} {\bibfnamefont {E.}~\bibnamefont {Waxman}},\ }\bibfield
  {journal} {\bibinfo  {journal} {arXiv e-prints}\ }\href
  {https://doi.org/10.48550/arXiv.1709.06507} {10.48550/arXiv.1709.06507}
  (\bibinfo {year} {2017}),\ \Eprint {https://arxiv.org/abs/1709.06507}
  {arxiv:1709.06507 [astro-ph.HE]} \BibitemShut {NoStop}%
\bibitem [{PAM()}]{PAMELAExperimentWeb}%
  \BibitemOpen
  \href {https://pamela-web.web.roma2.infn.it/} {\bibinfo {title} {{{PAMELA}}
  experiment web page}},\ \bibinfo {howpublished}
  {https://pamela-web.web.roma2.infn.it/}\BibitemShut {NoStop}%
\bibitem [{Fer()}]{FermiGammaraySpace}%
  \BibitemOpen
  \href {https://fermi.gsfc.nasa.gov/} {\bibinfo {title} {The {{Fermi Gamma-ray
  Space Telescope}}}},\ \bibinfo {howpublished}
  {https://fermi.gsfc.nasa.gov/}\BibitemShut {NoStop}%
\bibitem [{\citenamefont {Cui}\ \emph {et~al.}(2017)\citenamefont {Cui},
  \citenamefont {Yuan}, \citenamefont {Tsai},\ and\ \citenamefont
  {Fan}}]{Cui:2016ppb}%
  \BibitemOpen
  \bibfield  {author} {\bibinfo {author} {\bibfnamefont {M.-Y.}\ \bibnamefont
  {Cui}}, \bibinfo {author} {\bibfnamefont {Q.}~\bibnamefont {Yuan}}, \bibinfo
  {author} {\bibfnamefont {Y.-L.~S.}\ \bibnamefont {Tsai}},\ and\ \bibinfo
  {author} {\bibfnamefont {Y.-Z.}\ \bibnamefont {Fan}},\ }\href
  {https://doi.org/10.1103/PhysRevLett.118.191101} {\bibfield  {journal}
  {\bibinfo  {journal} {Physical Review Letters}\ }\textbf {\bibinfo {volume}
  {118}},\ \bibinfo {pages} {191101} (\bibinfo {year} {2017})},\ \Eprint
  {https://arxiv.org/abs/1610.03840} {arxiv:1610.03840 [astro-ph.HE]}
  \BibitemShut {NoStop}%
\bibitem [{\citenamefont {Cuoco}\ \emph
  {et~al.}(2017{\natexlab{a}})\citenamefont {Cuoco}, \citenamefont
  {Kr{\"a}mer},\ and\ \citenamefont {Korsmeier}}]{Cuoco:2016eej}%
  \BibitemOpen
  \bibfield  {author} {\bibinfo {author} {\bibfnamefont {A.}~\bibnamefont
  {Cuoco}}, \bibinfo {author} {\bibfnamefont {M.}~\bibnamefont {Kr{\"a}mer}},\
  and\ \bibinfo {author} {\bibfnamefont {M.}~\bibnamefont {Korsmeier}},\ }\href
  {https://doi.org/10.1103/PhysRevLett.118.191102} {\bibfield  {journal}
  {\bibinfo  {journal} {Physical Review Letters}\ }\textbf {\bibinfo {volume}
  {118}},\ \bibinfo {pages} {191102} (\bibinfo {year} {2017}{\natexlab{a}})},\
  \Eprint {https://arxiv.org/abs/1610.03071} {arxiv:1610.03071 [astro-ph.HE]}
  \BibitemShut {NoStop}%
\bibitem [{\citenamefont {Cui}\ \emph {et~al.}(2018)\citenamefont {Cui},
  \citenamefont {Pan}, \citenamefont {Yuan}, \citenamefont {Fan},\ and\
  \citenamefont {Zong}}]{Cui:2018klo}%
  \BibitemOpen
  \bibfield  {author} {\bibinfo {author} {\bibfnamefont {M.-Y.}\ \bibnamefont
  {Cui}}, \bibinfo {author} {\bibfnamefont {X.}~\bibnamefont {Pan}}, \bibinfo
  {author} {\bibfnamefont {Q.}~\bibnamefont {Yuan}}, \bibinfo {author}
  {\bibfnamefont {Y.-Z.}\ \bibnamefont {Fan}},\ and\ \bibinfo {author}
  {\bibfnamefont {H.-S.}\ \bibnamefont {Zong}},\ }\href
  {https://doi.org/10.1088/1475-7516/2018/06/024} {\bibfield  {journal}
  {\bibinfo  {journal} {Journal of Cosmology and Astroparticle Physics}\
  }\textbf {\bibinfo {volume} {06}}\bibfield  {number} {\bibinfo  {number} {
  (06)},\ \bibinfo {pages} {024}},\ }\Eprint {https://arxiv.org/abs/1803.02163}
  {arxiv:1803.02163 [astro-ph.HE]} \BibitemShut {NoStop}%
\bibitem [{\citenamefont {Cuoco}\ \emph {et~al.}(2019)\citenamefont {Cuoco},
  \citenamefont {Heisig}, \citenamefont {Klamt}, \citenamefont {Korsmeier},\
  and\ \citenamefont {Kr{\"a}mer}}]{Cuoco:2019kuu}%
  \BibitemOpen
  \bibfield  {author} {\bibinfo {author} {\bibfnamefont {A.}~\bibnamefont
  {Cuoco}}, \bibinfo {author} {\bibfnamefont {J.}~\bibnamefont {Heisig}},
  \bibinfo {author} {\bibfnamefont {L.}~\bibnamefont {Klamt}}, \bibinfo
  {author} {\bibfnamefont {M.}~\bibnamefont {Korsmeier}},\ and\ \bibinfo
  {author} {\bibfnamefont {M.}~\bibnamefont {Kr{\"a}mer}},\ }\href
  {https://doi.org/10.1103/PhysRevD.99.103014} {\bibfield  {journal} {\bibinfo
  {journal} {Physical Review D}\ }\textbf {\bibinfo {volume} {99}},\ \bibinfo
  {pages} {103014} (\bibinfo {year} {2019})},\ \Eprint
  {https://arxiv.org/abs/1903.01472} {arxiv:1903.01472 [astro-ph.HE]}
  \BibitemShut {NoStop}%
\bibitem [{\citenamefont {Cholis}\ \emph {et~al.}(2019)\citenamefont {Cholis},
  \citenamefont {Linden},\ and\ \citenamefont {Hooper}}]{Cholis:2019ejx}%
  \BibitemOpen
  \bibfield  {author} {\bibinfo {author} {\bibfnamefont {I.}~\bibnamefont
  {Cholis}}, \bibinfo {author} {\bibfnamefont {T.}~\bibnamefont {Linden}},\
  and\ \bibinfo {author} {\bibfnamefont {D.}~\bibnamefont {Hooper}},\ }\href
  {https://doi.org/10.1103/PhysRevD.99.103026} {\bibfield  {journal} {\bibinfo
  {journal} {Physical Review D}\ }\textbf {\bibinfo {volume} {99}},\ \bibinfo
  {pages} {103026} (\bibinfo {year} {2019})},\ \Eprint
  {https://arxiv.org/abs/1903.02549} {arxiv:1903.02549 [astro-ph.HE]}
  \BibitemShut {NoStop}%
\bibitem [{\citenamefont {Zhu}\ \emph {et~al.}(2022)\citenamefont {Zhu},
  \citenamefont {Cui}, \citenamefont {Xia}, \citenamefont {Yu}, \citenamefont
  {Huang}, \citenamefont {Yuan},\ and\ \citenamefont {Fan}}]{Zhu:2022tpr}%
  \BibitemOpen
  \bibfield  {author} {\bibinfo {author} {\bibfnamefont {C.-R.}\ \bibnamefont
  {Zhu}}, \bibinfo {author} {\bibfnamefont {M.-Y.}\ \bibnamefont {Cui}},
  \bibinfo {author} {\bibfnamefont {Z.-Q.}\ \bibnamefont {Xia}}, \bibinfo
  {author} {\bibfnamefont {Z.-H.}\ \bibnamefont {Yu}}, \bibinfo {author}
  {\bibfnamefont {X.}~\bibnamefont {Huang}}, \bibinfo {author} {\bibfnamefont
  {Q.}~\bibnamefont {Yuan}},\ and\ \bibinfo {author} {\bibfnamefont {Y.-Z.}\
  \bibnamefont {Fan}},\ }\href {https://doi.org/10.1103/PhysRevLett.129.231101}
  {\bibfield  {journal} {\bibinfo  {journal} {Phys. Rev. Lett.}\ }\textbf
  {\bibinfo {volume} {129}},\ \bibinfo {pages} {231101} (\bibinfo {year}
  {2022})},\ \Eprint {https://arxiv.org/abs/2204.03767} {arxiv:2204.03767
  [astro-ph.HE]} \BibitemShut {NoStop}%
\bibitem [{\citenamefont {Di~Mauro}\ and\ \citenamefont
  {Winkler}(2021)}]{DiMauro:2021qcf}%
  \BibitemOpen
  \bibfield  {author} {\bibinfo {author} {\bibfnamefont {M.}~\bibnamefont
  {Di~Mauro}}\ and\ \bibinfo {author} {\bibfnamefont {M.~W.}\ \bibnamefont
  {Winkler}},\ }\href {https://doi.org/10.1103/PhysRevD.103.123005} {\bibfield
  {journal} {\bibinfo  {journal} {Physical Review D}\ }\textbf {\bibinfo
  {volume} {103}},\ \bibinfo {pages} {123005} (\bibinfo {year} {2021})},\
  \Eprint {https://arxiv.org/abs/2101.11027} {arxiv:2101.11027 [astro-ph.HE]}
  \BibitemShut {NoStop}%
\bibitem [{\citenamefont {Cuoco}\ \emph
  {et~al.}(2017{\natexlab{b}})\citenamefont {Cuoco}, \citenamefont {Heisig},
  \citenamefont {Korsmeier},\ and\ \citenamefont {Kr{\"a}mer}}]{Cuoco:2017rxb}%
  \BibitemOpen
  \bibfield  {author} {\bibinfo {author} {\bibfnamefont {A.}~\bibnamefont
  {Cuoco}}, \bibinfo {author} {\bibfnamefont {J.}~\bibnamefont {Heisig}},
  \bibinfo {author} {\bibfnamefont {M.}~\bibnamefont {Korsmeier}},\ and\
  \bibinfo {author} {\bibfnamefont {M.}~\bibnamefont {Kr{\"a}mer}},\ }\href
  {https://doi.org/10.1088/1475-7516/2017/10/053} {\bibfield  {journal}
  {\bibinfo  {journal} {Journal of Cosmology and Astroparticle Physics}\
  }\textbf {\bibinfo {volume} {10}}\bibfield  {number} {\bibinfo  {number} {
  (10)},\ \bibinfo {pages} {053}},\ }\Eprint {https://arxiv.org/abs/1704.08258}
  {arxiv:1704.08258 [astro-ph.HE]} \BibitemShut {NoStop}%
\bibitem [{\citenamefont {Di~Mauro}(2021)}]{DiMauro:2021raz}%
  \BibitemOpen
  \bibfield  {author} {\bibinfo {author} {\bibfnamefont {M.}~\bibnamefont
  {Di~Mauro}},\ }\href {https://doi.org/10.1103/PhysRevD.103.063029} {\bibfield
   {journal} {\bibinfo  {journal} {Physical Review D}\ }\textbf {\bibinfo
  {volume} {103}},\ \bibinfo {pages} {063029} (\bibinfo {year} {2021})},\
  \Eprint {https://arxiv.org/abs/2101.04694} {arxiv:2101.04694 [astro-ph.HE]}
  \BibitemShut {NoStop}%
\bibitem [{\citenamefont {Leane}(2020)}]{Leane:2020liq}%
  \BibitemOpen
  \bibfield  {author} {\bibinfo {author} {\bibfnamefont {R.~K.}\ \bibnamefont
  {Leane}},\ }\href@noop {} {\bibfield  {journal} {\bibinfo  {journal} {arXiv
  e-prints}\ } (\bibinfo {year} {2020})},\ \Eprint
  {https://arxiv.org/abs/2006.00513} {arxiv:2006.00513 [hep-ph]} \BibitemShut
  {NoStop}%
\bibitem [{\citenamefont {Hooper}\ \emph {et~al.}(2020)\citenamefont {Hooper},
  \citenamefont {Leane}, \citenamefont {Tsai}, \citenamefont {Wegsman},\ and\
  \citenamefont {Witte}}]{Hooper:2019xss}%
  \BibitemOpen
  \bibfield  {author} {\bibinfo {author} {\bibfnamefont {D.}~\bibnamefont
  {Hooper}}, \bibinfo {author} {\bibfnamefont {R.~K.}\ \bibnamefont {Leane}},
  \bibinfo {author} {\bibfnamefont {Y.-D.}\ \bibnamefont {Tsai}}, \bibinfo
  {author} {\bibfnamefont {S.}~\bibnamefont {Wegsman}},\ and\ \bibinfo {author}
  {\bibfnamefont {S.~J.}\ \bibnamefont {Witte}},\ }\href
  {https://doi.org/10.1007/JHEP07(2020)163} {\bibfield  {journal} {\bibinfo
  {journal} {Journal of High Energy Physics}\ }\textbf {\bibinfo {volume}
  {07}},\ \bibinfo {pages} {163} (\bibinfo {year} {2020})},\ \Eprint
  {https://arxiv.org/abs/1912.08821} {arxiv:1912.08821 [hep-ph]} \BibitemShut
  {NoStop}%
\bibitem [{\citenamefont {Boudaud}\ \emph {et~al.}(2020)\citenamefont
  {Boudaud}, \citenamefont {G{\'e}nolini}, \citenamefont {Derome},
  \citenamefont {Lavalle}, \citenamefont {Maurin}, \citenamefont {Salati},\
  and\ \citenamefont {Serpico}}]{Boudaud:2019efq}%
  \BibitemOpen
  \bibfield  {author} {\bibinfo {author} {\bibfnamefont {M.}~\bibnamefont
  {Boudaud}}, \bibinfo {author} {\bibfnamefont {Y.}~\bibnamefont
  {G{\'e}nolini}}, \bibinfo {author} {\bibfnamefont {L.}~\bibnamefont
  {Derome}}, \bibinfo {author} {\bibfnamefont {J.}~\bibnamefont {Lavalle}},
  \bibinfo {author} {\bibfnamefont {D.}~\bibnamefont {Maurin}}, \bibinfo
  {author} {\bibfnamefont {P.}~\bibnamefont {Salati}},\ and\ \bibinfo {author}
  {\bibfnamefont {P.~D.}\ \bibnamefont {Serpico}},\ }\href
  {https://doi.org/10.1103/PhysRevResearch.2.023022} {\bibfield  {journal}
  {\bibinfo  {journal} {Physical Review Research}\ }\textbf {\bibinfo {volume}
  {2}},\ \bibinfo {pages} {023022} (\bibinfo {year} {2020})},\ \Eprint
  {https://arxiv.org/abs/1906.07119} {arxiv:1906.07119 [astro-ph.HE]}
  \BibitemShut {NoStop}%
\bibitem [{\citenamefont {Heisig}\ \emph {et~al.}(2020)\citenamefont {Heisig},
  \citenamefont {Korsmeier},\ and\ \citenamefont {Winkler}}]{Heisig:2020nse}%
  \BibitemOpen
  \bibfield  {author} {\bibinfo {author} {\bibfnamefont {J.}~\bibnamefont
  {Heisig}}, \bibinfo {author} {\bibfnamefont {M.}~\bibnamefont {Korsmeier}},\
  and\ \bibinfo {author} {\bibfnamefont {M.~W.}\ \bibnamefont {Winkler}},\
  }\href {https://doi.org/10.1103/PhysRevResearch.2.043017} {\bibfield
  {journal} {\bibinfo  {journal} {Physical Review Research}\ }\textbf {\bibinfo
  {volume} {2}},\ \bibinfo {pages} {043017} (\bibinfo {year} {2020})},\ \Eprint
  {https://arxiv.org/abs/2005.04237} {arxiv:2005.04237 [astro-ph.HE]}
  \BibitemShut {NoStop}%
\bibitem [{\citenamefont {Reinert}\ and\ \citenamefont
  {Winkler}(2018)}]{Reinert:2017aga}%
  \BibitemOpen
  \bibfield  {author} {\bibinfo {author} {\bibfnamefont {A.}~\bibnamefont
  {Reinert}}\ and\ \bibinfo {author} {\bibfnamefont {M.~W.}\ \bibnamefont
  {Winkler}},\ }\href {https://doi.org/10.1088/1475-7516/2018/01/055}
  {\bibfield  {journal} {\bibinfo  {journal} {JCAP}\ }\textbf {\bibinfo
  {volume} {2018}}\bibfield  {number} {\bibinfo  {number} { (01)},\ \bibinfo
  {pages} {055}},\ }\Eprint {https://arxiv.org/abs/1712.00002}
  {arxiv:1712.00002 [astro-ph.HE]} \BibitemShut {NoStop}%
\bibitem [{\citenamefont {Calore}\ \emph {et~al.}(2022)\citenamefont {Calore},
  \citenamefont {Cirelli}, \citenamefont {Derome}, \citenamefont {Genolini},
  \citenamefont {Maurin}, \citenamefont {Salati},\ and\ \citenamefont
  {Serpico}}]{caloreAMS02AntiprotonsDark2022}%
  \BibitemOpen
  \bibfield  {author} {\bibinfo {author} {\bibfnamefont {F.}~\bibnamefont
  {Calore}}, \bibinfo {author} {\bibfnamefont {M.}~\bibnamefont {Cirelli}},
  \bibinfo {author} {\bibfnamefont {L.}~\bibnamefont {Derome}}, \bibinfo
  {author} {\bibfnamefont {Y.}~\bibnamefont {Genolini}}, \bibinfo {author}
  {\bibfnamefont {D.}~\bibnamefont {Maurin}}, \bibinfo {author} {\bibfnamefont
  {P.}~\bibnamefont {Salati}},\ and\ \bibinfo {author} {\bibfnamefont {P.~D.}\
  \bibnamefont {Serpico}},\ }\href
  {https://doi.org/10.21468/SciPostPhys.12.5.163} {\bibfield  {journal}
  {\bibinfo  {journal} {SciPost Phys.}\ }\textbf {\bibinfo {volume} {12}},\
  \bibinfo {pages} {163} (\bibinfo {year} {2022})}\BibitemShut {NoStop}%
\bibitem [{\citenamefont {Lin}\ \emph {et~al.}(2017)\citenamefont {Lin},
  \citenamefont {Bi}, \citenamefont {Feng}, \citenamefont {Yin},\ and\
  \citenamefont {Yu}}]{Lin:2016ezz}%
  \BibitemOpen
  \bibfield  {author} {\bibinfo {author} {\bibfnamefont {S.-J.}\ \bibnamefont
  {Lin}}, \bibinfo {author} {\bibfnamefont {X.-J.}\ \bibnamefont {Bi}},
  \bibinfo {author} {\bibfnamefont {J.}~\bibnamefont {Feng}}, \bibinfo {author}
  {\bibfnamefont {P.-F.}\ \bibnamefont {Yin}},\ and\ \bibinfo {author}
  {\bibfnamefont {Z.-H.}\ \bibnamefont {Yu}},\ }\href
  {https://doi.org/10.1103/PhysRevD.96.123010} {\bibfield  {journal} {\bibinfo
  {journal} {Physical Review D}\ }\textbf {\bibinfo {volume} {96}},\ \bibinfo
  {pages} {123010} (\bibinfo {year} {2017})},\ \Eprint
  {https://arxiv.org/abs/1612.04001} {arxiv:1612.04001 [astro-ph.HE]}
  \BibitemShut {NoStop}%
\bibitem [{\citenamefont {Lin}\ \emph {et~al.}(2019)\citenamefont {Lin},
  \citenamefont {Bi},\ and\ \citenamefont {Yin}}]{Lin:2019ljc}%
  \BibitemOpen
  \bibfield  {author} {\bibinfo {author} {\bibfnamefont {S.-J.}\ \bibnamefont
  {Lin}}, \bibinfo {author} {\bibfnamefont {X.-J.}\ \bibnamefont {Bi}},\ and\
  \bibinfo {author} {\bibfnamefont {P.-F.}\ \bibnamefont {Yin}},\ }\href
  {https://doi.org/10.1103/PhysRevD.100.103014} {\bibfield  {journal} {\bibinfo
   {journal} {Physical Review D}\ }\textbf {\bibinfo {volume} {100}},\ \bibinfo
  {pages} {103014} (\bibinfo {year} {2019})},\ \Eprint
  {https://arxiv.org/abs/1903.09545} {arxiv:1903.09545 [astro-ph.HE]}
  \BibitemShut {NoStop}%
\bibitem [{\citenamefont {Luque}(2021)}]{Luque:2021ddh}%
  \BibitemOpen
  \bibfield  {author} {\bibinfo {author} {\bibfnamefont {P.~D. L.~T.}\
  \bibnamefont {Luque}},\ }\href
  {https://doi.org/10.1088/1475-7516/2021/11/018} {\bibfield  {journal}
  {\bibinfo  {journal} {JCAP}\ }\textbf {\bibinfo {volume} {2021}}\bibfield
  {number} {\bibinfo  {number} { (11)},\ \bibinfo {pages} {018}},\ }\Eprint
  {https://arxiv.org/abs/2107.06863} {arxiv:2107.06863 [astro-ph.HE]}
  \BibitemShut {NoStop}%
\bibitem [{\citenamefont {Donato}\ \emph {et~al.}(2001)\citenamefont {Donato},
  \citenamefont {Maurin}, \citenamefont {Salati}, \citenamefont {Barrau},
  \citenamefont {Boudoul},\ and\ \citenamefont {Taillet}}]{Donato:2001ms}%
  \BibitemOpen
  \bibfield  {author} {\bibinfo {author} {\bibfnamefont {F.}~\bibnamefont
  {Donato}}, \bibinfo {author} {\bibfnamefont {D.}~\bibnamefont {Maurin}},
  \bibinfo {author} {\bibfnamefont {P.}~\bibnamefont {Salati}}, \bibinfo
  {author} {\bibfnamefont {A.}~\bibnamefont {Barrau}}, \bibinfo {author}
  {\bibfnamefont {G.}~\bibnamefont {Boudoul}},\ and\ \bibinfo {author}
  {\bibfnamefont {R.}~\bibnamefont {Taillet}},\ }\href
  {https://doi.org/10.1086/323684} {\bibfield  {journal} {\bibinfo  {journal}
  {Astrophys. J.}\ }\textbf {\bibinfo {volume} {563}},\ \bibinfo {pages} {172}
  (\bibinfo {year} {2001})},\ \Eprint {https://arxiv.org/abs/astro-ph/0103150}
  {arxiv:astro-ph/0103150} \BibitemShut {NoStop}%
\bibitem [{\citenamefont {Yuan}\ \emph {et~al.}(2017)\citenamefont {Yuan},
  \citenamefont {Lin}, \citenamefont {Fang},\ and\ \citenamefont
  {Bi}}]{Yuan:2017ozr}%
  \BibitemOpen
  \bibfield  {author} {\bibinfo {author} {\bibfnamefont {Q.}~\bibnamefont
  {Yuan}}, \bibinfo {author} {\bibfnamefont {S.-J.}\ \bibnamefont {Lin}},
  \bibinfo {author} {\bibfnamefont {K.}~\bibnamefont {Fang}},\ and\ \bibinfo
  {author} {\bibfnamefont {X.-J.}\ \bibnamefont {Bi}},\ }\href
  {https://doi.org/10.1103/PhysRevD.95.083007} {\bibfield  {journal} {\bibinfo
  {journal} {Phys. Rev. D}\ }\textbf {\bibinfo {volume} {95}},\ \bibinfo
  {pages} {083007} (\bibinfo {year} {2017})},\ \Eprint
  {https://arxiv.org/abs/1701.06149} {arXiv:1701.06149 [astro-ph.HE]}
  \BibitemShut {NoStop}%
\bibitem [{\citenamefont {Yuan}(2018)}]{Yuan:2018vgk}%
  \BibitemOpen
  \bibfield  {author} {\bibinfo {author} {\bibfnamefont {Q.}~\bibnamefont
  {Yuan}},\ }\href {https://doi.org/10.1007/s11433-018-9300-0} {\bibfield
  {journal} {\bibinfo  {journal} {Science China Physics, Mechanics \&
  Astronomy}\ }\textbf {\bibinfo {volume} {62}},\ \bibinfo {pages} {49511}
  (\bibinfo {year} {2018})},\ \Eprint {https://arxiv.org/abs/1805.10649}
  {arxiv:1805.10649 [astro-ph.HE]} \BibitemShut {NoStop}%
\bibitem [{\citenamefont
  {Tomassetti}(2017)}]{tomassettiSolarNuclearPhysics2017}%
  \BibitemOpen
  \bibfield  {author} {\bibinfo {author} {\bibfnamefont {N.}~\bibnamefont
  {Tomassetti}},\ }\href {https://doi.org/10.1103/PhysRevD.96.103005}
  {\bibfield  {journal} {\bibinfo  {journal} {Physical Review D}\ }\textbf
  {\bibinfo {volume} {96}},\ \bibinfo {pages} {103005} (\bibinfo {year}
  {2017})},\ \Eprint {https://arxiv.org/abs/1707.06917} {arxiv:1707.06917
  [astro-ph, physics:hep-ph]} \BibitemShut {NoStop}%
\bibitem [{\citenamefont {G{\'e}nolini}\ \emph {et~al.}(2018)\citenamefont
  {G{\'e}nolini}, \citenamefont {Maurin}, \citenamefont {Moskalenko},\ and\
  \citenamefont {Unger}}]{Genolini:2018ekk}%
  \BibitemOpen
  \bibfield  {author} {\bibinfo {author} {\bibfnamefont {Y.}~\bibnamefont
  {G{\'e}nolini}}, \bibinfo {author} {\bibfnamefont {D.}~\bibnamefont
  {Maurin}}, \bibinfo {author} {\bibfnamefont {I.~V.}\ \bibnamefont
  {Moskalenko}},\ and\ \bibinfo {author} {\bibfnamefont {M.}~\bibnamefont
  {Unger}},\ }\href {https://doi.org/10.1103/PhysRevC.98.034611} {\bibfield
  {journal} {\bibinfo  {journal} {Physical Review C}\ }\textbf {\bibinfo
  {volume} {98}},\ \bibinfo {pages} {034611} (\bibinfo {year} {2018})},\
  \Eprint {https://arxiv.org/abs/1803.04686} {arxiv:1803.04686 [astro-ph.HE]}
  \BibitemShut {NoStop}%
\bibitem [{\citenamefont {Weinrich}\ \emph
  {et~al.}(2020{\natexlab{a}})\citenamefont {Weinrich}, \citenamefont
  {G{\'e}nolini}, \citenamefont {Boudaud}, \citenamefont {Derome},\ and\
  \citenamefont {Maurin}}]{Weinrich:2020cmw}%
  \BibitemOpen
  \bibfield  {author} {\bibinfo {author} {\bibfnamefont {N.}~\bibnamefont
  {Weinrich}}, \bibinfo {author} {\bibfnamefont {Y.}~\bibnamefont
  {G{\'e}nolini}}, \bibinfo {author} {\bibfnamefont {M.}~\bibnamefont
  {Boudaud}}, \bibinfo {author} {\bibfnamefont {L.}~\bibnamefont {Derome}},\
  and\ \bibinfo {author} {\bibfnamefont {D.}~\bibnamefont {Maurin}},\ }\href
  {https://doi.org/10.1051/0004-6361/202037875} {\bibfield  {journal} {\bibinfo
   {journal} {Astronomy \& Astrophysics}\ }\textbf {\bibinfo {volume} {639}},\
  \bibinfo {pages} {A131} (\bibinfo {year} {2020}{\natexlab{a}})},\ \Eprint
  {https://arxiv.org/abs/2002.11406} {arxiv:2002.11406 [astro-ph.HE]}
  \BibitemShut {NoStop}%
\bibitem [{\citenamefont {Korsmeier}\ and\ \citenamefont
  {Cuoco}(2021)}]{Korsmeier:2021brc}%
  \BibitemOpen
  \bibfield  {author} {\bibinfo {author} {\bibfnamefont {M.}~\bibnamefont
  {Korsmeier}}\ and\ \bibinfo {author} {\bibfnamefont {A.}~\bibnamefont
  {Cuoco}},\ }\href {https://doi.org/10.1103/PhysRevD.103.103016} {\bibfield
  {journal} {\bibinfo  {journal} {Physical Review D}\ }\textbf {\bibinfo
  {volume} {103}},\ \bibinfo {pages} {103016} (\bibinfo {year} {2021})},\
  \Eprint {https://arxiv.org/abs/2103.09824} {arxiv:2103.09824 [astro-ph.HE]}
  \BibitemShut {NoStop}%
\bibitem [{\citenamefont {Potgieter}(2013)}]{Potgieter:2013pdj}%
  \BibitemOpen
  \bibfield  {author} {\bibinfo {author} {\bibfnamefont {M.~S.}\ \bibnamefont
  {Potgieter}},\ }\href {https://doi.org/10.12942/lrsp-2013-3} {\bibfield
  {journal} {\bibinfo  {journal} {Living Reviews in Solar Physics}\ }\textbf
  {\bibinfo {volume} {10}},\ \bibinfo {pages} {3} (\bibinfo {year} {2013})},\
  \Eprint {https://arxiv.org/abs/1306.4421} {arxiv:1306.4421
  [physics.space-ph]} \BibitemShut {NoStop}%
\bibitem [{\citenamefont {Strong}\ and\ \citenamefont
  {Moskalenko}(1998)}]{Strong:1998pw}%
  \BibitemOpen
  \bibfield  {author} {\bibinfo {author} {\bibfnamefont {A.~W.}\ \bibnamefont
  {Strong}}\ and\ \bibinfo {author} {\bibfnamefont {I.~V.}\ \bibnamefont
  {Moskalenko}},\ }\href {https://doi.org/10.1086/306470} {\bibfield  {journal}
  {\bibinfo  {journal} {The Astrophysical Journal}\ }\textbf {\bibinfo {volume}
  {509}},\ \bibinfo {pages} {212} (\bibinfo {year} {1998})},\ \Eprint
  {https://arxiv.org/abs/astro-ph/9807150} {arxiv:astro-ph/9807150}
  \BibitemShut {NoStop}%
\bibitem [{\citenamefont {Strong}\ \emph {et~al.}(2000)\citenamefont {Strong},
  \citenamefont {Moskalenko},\ and\ \citenamefont {Reimer}}]{Strong:1998fr}%
  \BibitemOpen
  \bibfield  {author} {\bibinfo {author} {\bibfnamefont {A.~W.}\ \bibnamefont
  {Strong}}, \bibinfo {author} {\bibfnamefont {I.~V.}\ \bibnamefont
  {Moskalenko}},\ and\ \bibinfo {author} {\bibfnamefont {O.}~\bibnamefont
  {Reimer}},\ }\href {https://doi.org/10.1086/309038} {\bibfield  {journal}
  {\bibinfo  {journal} {The Astrophysical Journal}\ }\textbf {\bibinfo {volume}
  {537}},\ \bibinfo {pages} {763} (\bibinfo {year} {2000})},\ \Eprint
  {https://arxiv.org/abs/astro-ph/9811296} {arxiv:astro-ph/9811296}
  \BibitemShut {NoStop}%
\bibitem [{\citenamefont {Lewis}\ and\ \citenamefont
  {Bridle}(2002)}]{Lewis:2002ah}%
  \BibitemOpen
  \bibfield  {author} {\bibinfo {author} {\bibfnamefont {A.}~\bibnamefont
  {Lewis}}\ and\ \bibinfo {author} {\bibfnamefont {S.}~\bibnamefont {Bridle}},\
  }\href {https://doi.org/10.1103/PhysRevD.66.103511} {\bibfield  {journal}
  {\bibinfo  {journal} {Phys. Rev.}\ }\textbf {\bibinfo {volume} {D66}},\
  \bibinfo {pages} {103511} (\bibinfo {year} {2002})},\ \Eprint
  {https://arxiv.org/abs/astro-ph/0205436} {arXiv:astro-ph/0205436 [astro-ph]}
  \BibitemShut {NoStop}%
\bibitem [{\citenamefont {Cholis}\ \emph {et~al.}(2016)\citenamefont {Cholis},
  \citenamefont {Hooper},\ and\ \citenamefont {Linden}}]{Cholis:2015gna}%
  \BibitemOpen
  \bibfield  {author} {\bibinfo {author} {\bibfnamefont {I.}~\bibnamefont
  {Cholis}}, \bibinfo {author} {\bibfnamefont {D.}~\bibnamefont {Hooper}},\
  and\ \bibinfo {author} {\bibfnamefont {T.}~\bibnamefont {Linden}},\ }\href
  {https://doi.org/10.1103/PhysRevD.93.043016} {\bibfield  {journal} {\bibinfo
  {journal} {Physical Review D}\ }\textbf {\bibinfo {volume} {93}},\ \bibinfo
  {pages} {043016} (\bibinfo {year} {2016})},\ \Eprint
  {https://arxiv.org/abs/1511.01507} {arxiv:1511.01507 [astro-ph.SR]}
  \BibitemShut {NoStop}%
\bibitem [{\citenamefont {Baatar}\ \emph {et~al.}(2013)\citenamefont {Baatar},
  \citenamefont {Barr}, \citenamefont {Bartke} \emph {et~al.}}]{NA49:2012jna}%
  \BibitemOpen
  \bibfield  {author} {\bibinfo {author} {\bibfnamefont {B.}~\bibnamefont
  {Baatar}}, \bibinfo {author} {\bibfnamefont {G.}~\bibnamefont {Barr}},
  \bibinfo {author} {\bibfnamefont {J.}~\bibnamefont {Bartke}}, \emph {et~al.}
  (\bibinfo {collaboration} {NA49}),\ }\href
  {https://doi.org/10.1140/epjc/s10052-013-2364-3} {\bibfield  {journal}
  {\bibinfo  {journal} {Eur. Phys. J. C}\ }\textbf {\bibinfo {volume} {73}},\
  \bibinfo {pages} {2364} (\bibinfo {year} {2013})},\ \Eprint
  {https://arxiv.org/abs/1207.6520} {arxiv:1207.6520 [hep-ex]} \BibitemShut
  {NoStop}%
\bibitem [{\citenamefont {Collaboration}(2017)}]{NA61SHINE:2017fne}%
  \BibitemOpen
  \bibfield  {author} {\bibinfo {author} {\bibfnamefont {N.}~\bibnamefont
  {Collaboration}} (\bibinfo {collaboration} {NA61/SHINE}),\ }\href
  {https://doi.org/10.1140/epjc/s10052-017-5260-4} {\bibfield  {journal}
  {\bibinfo  {journal} {The European Physical Journal C}\ }\textbf {\bibinfo
  {volume} {77}},\ \bibinfo {pages} {671} (\bibinfo {year} {2017})},\ \Eprint
  {https://arxiv.org/abs/1705.02467} {arxiv:1705.02467 [nucl-ex]} \BibitemShut
  {NoStop}%
\bibitem [{\citenamefont {Aaij}\ \emph {et~al.}(2018)\citenamefont {Aaij} \emph
  {et~al.}}]{LHCb:2018ygc}%
  \BibitemOpen
  \bibfield  {author} {\bibinfo {author} {\bibfnamefont {R.}~\bibnamefont
  {Aaij}} \emph {et~al.} (\bibinfo {collaboration} {LHCb}),\ }\href
  {https://doi.org/10.1103/PhysRevLett.121.222001} {\bibfield  {journal}
  {\bibinfo  {journal} {Phys. Rev. Lett.}\ }\textbf {\bibinfo {volume} {121}},\
  \bibinfo {pages} {222001} (\bibinfo {year} {2018})},\ \Eprint
  {https://arxiv.org/abs/1808.06127} {arxiv:1808.06127 [hep-ex]} \BibitemShut
  {NoStop}%
\bibitem [{\citenamefont {Korsmeier}\ \emph {et~al.}(2018)\citenamefont
  {Korsmeier}, \citenamefont {Donato},\ and\ \citenamefont
  {Di~Mauro}}]{Korsmeier:2018gcy}%
  \BibitemOpen
  \bibfield  {author} {\bibinfo {author} {\bibfnamefont {M.}~\bibnamefont
  {Korsmeier}}, \bibinfo {author} {\bibfnamefont {F.}~\bibnamefont {Donato}},\
  and\ \bibinfo {author} {\bibfnamefont {M.}~\bibnamefont {Di~Mauro}},\ }\href
  {https://doi.org/10.1103/PhysRevD.97.103019} {\bibfield  {journal} {\bibinfo
  {journal} {Physical Review D}\ }\textbf {\bibinfo {volume} {97}},\ \bibinfo
  {pages} {103019} (\bibinfo {year} {2018})},\ \Eprint
  {https://arxiv.org/abs/1802.03030} {arxiv:1802.03030 [astro-ph.HE]}
  \BibitemShut {NoStop}%
\bibitem [{\citenamefont {Strong}\ \emph {et~al.}(2007)\citenamefont {Strong},
  \citenamefont {Moskalenko},\ and\ \citenamefont {Ptuskin}}]{Strong:2007nh}%
  \BibitemOpen
  \bibfield  {author} {\bibinfo {author} {\bibfnamefont {A.~W.}\ \bibnamefont
  {Strong}}, \bibinfo {author} {\bibfnamefont {I.~V.}\ \bibnamefont
  {Moskalenko}},\ and\ \bibinfo {author} {\bibfnamefont {V.~S.}\ \bibnamefont
  {Ptuskin}},\ }\href {https://doi.org/10.1146/annurev.nucl.57.090506.123011}
  {\bibfield  {journal} {\bibinfo  {journal} {Annual Review of Nuclear and
  Particle Science}\ }\textbf {\bibinfo {volume} {57}},\ \bibinfo {pages} {285}
  (\bibinfo {year} {2007})},\ \Eprint {https://arxiv.org/abs/astro-ph/0701517}
  {arxiv:astro-ph/0701517} \BibitemShut {NoStop}%
\bibitem [{\citenamefont {Trotta}\ \emph {et~al.}(2011)\citenamefont {Trotta},
  \citenamefont {Johannesson}, \citenamefont {Moskalenko}, \citenamefont
  {Porter}, \citenamefont {{de Austri}},\ and\ \citenamefont
  {Strong}}]{Trotta:2010mx}%
  \BibitemOpen
  \bibfield  {author} {\bibinfo {author} {\bibfnamefont {R.}~\bibnamefont
  {Trotta}}, \bibinfo {author} {\bibfnamefont {G.}~\bibnamefont {Johannesson}},
  \bibinfo {author} {\bibfnamefont {I.~V.}\ \bibnamefont {Moskalenko}},
  \bibinfo {author} {\bibfnamefont {T.~A.}\ \bibnamefont {Porter}}, \bibinfo
  {author} {\bibfnamefont {R.~R.}\ \bibnamefont {{de Austri}}},\ and\ \bibinfo
  {author} {\bibfnamefont {A.~W.}\ \bibnamefont {Strong}},\ }\href
  {https://doi.org/10.1088/0004-637X/729/2/106} {\bibfield  {journal} {\bibinfo
   {journal} {The Astrophysical Journal}\ }\textbf {\bibinfo {volume} {729}},\
  \bibinfo {pages} {106} (\bibinfo {year} {2011})},\ \Eprint
  {https://arxiv.org/abs/1011.0037} {arxiv:1011.0037 [astro-ph.HE]}
  \BibitemShut {NoStop}%
\bibitem [{\citenamefont {Achterberg}\ \emph {et~al.}(2001)\citenamefont
  {Achterberg}, \citenamefont {Gallant}, \citenamefont {Kirk},\ and\
  \citenamefont {Guthmann}}]{Achterberg:2001rx}%
  \BibitemOpen
  \bibfield  {author} {\bibinfo {author} {\bibfnamefont {A.}~\bibnamefont
  {Achterberg}}, \bibinfo {author} {\bibfnamefont {Y.~A.}\ \bibnamefont
  {Gallant}}, \bibinfo {author} {\bibfnamefont {J.~G.}\ \bibnamefont {Kirk}},\
  and\ \bibinfo {author} {\bibfnamefont {A.~W.}\ \bibnamefont {Guthmann}},\
  }\href {https://doi.org/10.1046/j.1365-8711.2001.04851.x} {\bibfield
  {journal} {\bibinfo  {journal} {Monthly Notices of the Royal Astronomical
  Society}\ }\textbf {\bibinfo {volume} {328}},\ \bibinfo {pages} {393}
  (\bibinfo {year} {2001})},\ \Eprint {https://arxiv.org/abs/astro-ph/0107530}
  {arxiv:astro-ph/0107530} \BibitemShut {NoStop}%
\bibitem [{\citenamefont {Ahn}\ \emph {et~al.}(2010)\citenamefont {Ahn},
  \citenamefont {Allison}, \citenamefont {Bagliesi}, \citenamefont {Beatty},
  \citenamefont {Bigongiari}, \citenamefont {Childers}, \citenamefont
  {Conklin}, \citenamefont {Coutu}, \citenamefont {DuVernois}, \citenamefont
  {Ganel}, \citenamefont {Han}, \citenamefont {Jeon}, \citenamefont {Kim},
  \citenamefont {Lee}, \citenamefont {Lutz}, \citenamefont {Maestro},
  \citenamefont {Malinin}, \citenamefont {Marrocchesi}, \citenamefont
  {Minnick}, \citenamefont {Mognet}, \citenamefont {Nam}, \citenamefont {Nam},
  \citenamefont {Nutter}, \citenamefont {Park}, \citenamefont {Park},
  \citenamefont {Seo}, \citenamefont {Sina}, \citenamefont {Wu}, \citenamefont
  {Yang}, \citenamefont {Yoon}, \citenamefont {Zei},\ and\ \citenamefont
  {Zinn}}]{Ahn:2010gv}%
  \BibitemOpen
  \bibfield  {author} {\bibinfo {author} {\bibfnamefont {H.~S.}\ \bibnamefont
  {Ahn}}, \bibinfo {author} {\bibfnamefont {P.}~\bibnamefont {Allison}},
  \bibinfo {author} {\bibfnamefont {M.~G.}\ \bibnamefont {Bagliesi}}, \bibinfo
  {author} {\bibfnamefont {J.~J.}\ \bibnamefont {Beatty}}, \bibinfo {author}
  {\bibfnamefont {G.}~\bibnamefont {Bigongiari}}, \bibinfo {author}
  {\bibfnamefont {J.~T.}\ \bibnamefont {Childers}}, \bibinfo {author}
  {\bibfnamefont {N.~B.}\ \bibnamefont {Conklin}}, \bibinfo {author}
  {\bibfnamefont {S.}~\bibnamefont {Coutu}}, \bibinfo {author} {\bibfnamefont
  {M.~A.}\ \bibnamefont {DuVernois}}, \bibinfo {author} {\bibfnamefont
  {O.}~\bibnamefont {Ganel}}, \bibinfo {author} {\bibfnamefont {J.~H.}\
  \bibnamefont {Han}}, \bibinfo {author} {\bibfnamefont {J.~A.}\ \bibnamefont
  {Jeon}}, \bibinfo {author} {\bibfnamefont {K.~C.}\ \bibnamefont {Kim}},
  \bibinfo {author} {\bibfnamefont {M.~H.}\ \bibnamefont {Lee}}, \bibinfo
  {author} {\bibfnamefont {L.}~\bibnamefont {Lutz}}, \bibinfo {author}
  {\bibfnamefont {P.}~\bibnamefont {Maestro}}, \bibinfo {author} {\bibfnamefont
  {A.}~\bibnamefont {Malinin}}, \bibinfo {author} {\bibfnamefont {P.~S.}\
  \bibnamefont {Marrocchesi}}, \bibinfo {author} {\bibfnamefont
  {S.}~\bibnamefont {Minnick}}, \bibinfo {author} {\bibfnamefont {S.~I.}\
  \bibnamefont {Mognet}}, \bibinfo {author} {\bibfnamefont {J.}~\bibnamefont
  {Nam}}, \bibinfo {author} {\bibfnamefont {S.}~\bibnamefont {Nam}}, \bibinfo
  {author} {\bibfnamefont {S.~L.}\ \bibnamefont {Nutter}}, \bibinfo {author}
  {\bibfnamefont {I.~H.}\ \bibnamefont {Park}}, \bibinfo {author}
  {\bibfnamefont {N.~H.}\ \bibnamefont {Park}}, \bibinfo {author}
  {\bibfnamefont {E.~S.}\ \bibnamefont {Seo}}, \bibinfo {author} {\bibfnamefont
  {R.}~\bibnamefont {Sina}}, \bibinfo {author} {\bibfnamefont {J.}~\bibnamefont
  {Wu}}, \bibinfo {author} {\bibfnamefont {J.}~\bibnamefont {Yang}}, \bibinfo
  {author} {\bibfnamefont {Y.~S.}\ \bibnamefont {Yoon}}, \bibinfo {author}
  {\bibfnamefont {R.}~\bibnamefont {Zei}},\ and\ \bibinfo {author}
  {\bibfnamefont {S.~Y.}\ \bibnamefont {Zinn}},\ }\href
  {https://doi.org/10.1088/2041-8205/714/1/L89} {\bibfield  {journal} {\bibinfo
   {journal} {The Astrophysical Journal}\ }\textbf {\bibinfo {volume} {714}},\
  \bibinfo {pages} {L89} (\bibinfo {year} {2010})},\ \Eprint
  {https://arxiv.org/abs/1004.1123} {arxiv:1004.1123 [astro-ph.HE]}
  \BibitemShut {NoStop}%
\bibitem [{\citenamefont {Maurin}\ \emph {et~al.}(2010)\citenamefont {Maurin},
  \citenamefont {Putze},\ and\ \citenamefont {Derome}}]{Maurin:2010zp}%
  \BibitemOpen
  \bibfield  {author} {\bibinfo {author} {\bibfnamefont {D.}~\bibnamefont
  {Maurin}}, \bibinfo {author} {\bibfnamefont {A.}~\bibnamefont {Putze}},\ and\
  \bibinfo {author} {\bibfnamefont {L.}~\bibnamefont {Derome}},\ }\href
  {https://doi.org/10.1051/0004-6361/201014011} {\bibfield  {journal} {\bibinfo
   {journal} {Astronomy and Astrophysics}\ }\textbf {\bibinfo {volume} {516}},\
  \bibinfo {pages} {A67} (\bibinfo {year} {2010})},\ \Eprint
  {https://arxiv.org/abs/1001.0553} {arxiv:1001.0553 [astro-ph.HE]}
  \BibitemShut {NoStop}%
\bibitem [{\citenamefont {Drury}\ and\ \citenamefont
  {Strong}(2017)}]{Drury:2016ubm}%
  \BibitemOpen
  \bibfield  {author} {\bibinfo {author} {\bibfnamefont {L.~O.}\ \bibnamefont
  {Drury}}\ and\ \bibinfo {author} {\bibfnamefont {A.~W.}\ \bibnamefont
  {Strong}},\ }\href {https://doi.org/10.1051/0004-6361/201629526} {\bibfield
  {journal} {\bibinfo  {journal} {Astronomy \& Astrophysics}\ }\textbf
  {\bibinfo {volume} {597}},\ \bibinfo {pages} {A117} (\bibinfo {year}
  {2017})},\ \Eprint {https://arxiv.org/abs/1608.04227} {arxiv:1608.04227
  [astro-ph.HE]} \BibitemShut {NoStop}%
\bibitem [{\citenamefont {Gleeson}\ and\ \citenamefont
  {Axford}(1968)}]{Gleeson:1968zza}%
  \BibitemOpen
  \bibfield  {author} {\bibinfo {author} {\bibfnamefont {L.~J.}\ \bibnamefont
  {Gleeson}}\ and\ \bibinfo {author} {\bibfnamefont {W.~I.}\ \bibnamefont
  {Axford}},\ }\href {https://doi.org/10.1086/149822} {\bibfield  {journal}
  {\bibinfo  {journal} {The Astrophysical Journal}\ }\textbf {\bibinfo {volume}
  {154}},\ \bibinfo {pages} {1011} (\bibinfo {year} {1968})}\BibitemShut
  {NoStop}%
\bibitem [{\citenamefont {Strauss}\ \emph {et~al.}(2012)\citenamefont
  {Strauss}, \citenamefont {Potgieter}, \citenamefont {B{\"u}sching},\ and\
  \citenamefont {Kopp}}]{straussModellingHeliosphericCurrent2012}%
  \BibitemOpen
  \bibfield  {author} {\bibinfo {author} {\bibfnamefont {R.~D.}\ \bibnamefont
  {Strauss}}, \bibinfo {author} {\bibfnamefont {M.~S.}\ \bibnamefont
  {Potgieter}}, \bibinfo {author} {\bibfnamefont {I.}~\bibnamefont
  {B{\"u}sching}},\ and\ \bibinfo {author} {\bibfnamefont {A.}~\bibnamefont
  {Kopp}},\ }\href {https://doi.org/10.1007/s10509-012-1003-z} {\bibfield
  {journal} {\bibinfo  {journal} {Astrophysics and Space Science}\ }\textbf
  {\bibinfo {volume} {339}},\ \bibinfo {pages} {223} (\bibinfo {year}
  {2012})}\BibitemShut {NoStop}%
\bibitem [{\citenamefont {Weinrich}\ \emph
  {et~al.}(2020{\natexlab{b}})\citenamefont {Weinrich}, \citenamefont
  {Boudaud}, \citenamefont {Derome}, \citenamefont {G{\'e}nolini},
  \citenamefont {Lavalle}, \citenamefont {Maurin}, \citenamefont {Salati},
  \citenamefont {Serpico},\ and\ \citenamefont
  {{Weymann-Despres}}}]{Weinrich:2020ftb}%
  \BibitemOpen
  \bibfield  {author} {\bibinfo {author} {\bibfnamefont {N.}~\bibnamefont
  {Weinrich}}, \bibinfo {author} {\bibfnamefont {M.}~\bibnamefont {Boudaud}},
  \bibinfo {author} {\bibfnamefont {L.}~\bibnamefont {Derome}}, \bibinfo
  {author} {\bibfnamefont {Y.}~\bibnamefont {G{\'e}nolini}}, \bibinfo {author}
  {\bibfnamefont {J.}~\bibnamefont {Lavalle}}, \bibinfo {author} {\bibfnamefont
  {D.}~\bibnamefont {Maurin}}, \bibinfo {author} {\bibfnamefont
  {P.}~\bibnamefont {Salati}}, \bibinfo {author} {\bibfnamefont
  {P.}~\bibnamefont {Serpico}},\ and\ \bibinfo {author} {\bibfnamefont
  {G.}~\bibnamefont {{Weymann-Despres}}},\ }\href
  {https://doi.org/10.1051/0004-6361/202038064} {\bibfield  {journal} {\bibinfo
   {journal} {Astronomy \& Astrophysics}\ }\textbf {\bibinfo {volume} {639}},\
  \bibinfo {pages} {A74} (\bibinfo {year} {2020}{\natexlab{b}})},\ \Eprint
  {https://arxiv.org/abs/2004.00441} {arxiv:2004.00441 [astro-ph.HE]}
  \BibitemShut {NoStop}%
\bibitem [{\citenamefont {Luque}\ \emph {et~al.}(2021)\citenamefont {Luque},
  \citenamefont {Mazziotta}, \citenamefont {Loparco}, \citenamefont {Gargano},\
  and\ \citenamefont {Serini}}]{DeLaTorreLuque:2021yfq}%
  \BibitemOpen
  \bibfield  {author} {\bibinfo {author} {\bibfnamefont {P.~d. l.~T.}\
  \bibnamefont {Luque}}, \bibinfo {author} {\bibfnamefont {M.~N.}\ \bibnamefont
  {Mazziotta}}, \bibinfo {author} {\bibfnamefont {F.}~\bibnamefont {Loparco}},
  \bibinfo {author} {\bibfnamefont {F.}~\bibnamefont {Gargano}},\ and\ \bibinfo
  {author} {\bibfnamefont {D.}~\bibnamefont {Serini}},\ }\href
  {https://doi.org/10.1088/1475-7516/2021/03/099} {\bibfield  {journal}
  {\bibinfo  {journal} {Journal of Cosmology and Astroparticle Physics}\
  }\textbf {\bibinfo {volume} {03}}\bibfield  {number} {\bibinfo  {number} {
  (03)},\ \bibinfo {pages} {099}},\ }\Eprint {https://arxiv.org/abs/2101.01547}
  {arxiv:2101.01547 [astro-ph.HE]} \BibitemShut {NoStop}%
\bibitem [{\citenamefont {Evoli}\ \emph {et~al.}(2020)\citenamefont {Evoli},
  \citenamefont {Morlino}, \citenamefont {Blasi},\ and\ \citenamefont
  {Aloisio}}]{Evoli:2019iih}%
  \BibitemOpen
  \bibfield  {author} {\bibinfo {author} {\bibfnamefont {C.}~\bibnamefont
  {Evoli}}, \bibinfo {author} {\bibfnamefont {G.}~\bibnamefont {Morlino}},
  \bibinfo {author} {\bibfnamefont {P.}~\bibnamefont {Blasi}},\ and\ \bibinfo
  {author} {\bibfnamefont {R.}~\bibnamefont {Aloisio}},\ }\href
  {https://doi.org/10.1103/PhysRevD.101.023013} {\bibfield  {journal} {\bibinfo
   {journal} {Physical Review D}\ }\textbf {\bibinfo {volume} {101}},\ \bibinfo
  {pages} {023013} (\bibinfo {year} {2020})},\ \Eprint
  {https://arxiv.org/abs/1910.04113} {arxiv:1910.04113 [astro-ph.HE]}
  \BibitemShut {NoStop}%
\bibitem [{\citenamefont {Zhao}\ \emph {et~al.}(2023)\citenamefont {Zhao},
  \citenamefont {Bi},\ and\ \citenamefont {Fang}}]{Zhao:2022bon}%
  \BibitemOpen
  \bibfield  {author} {\bibinfo {author} {\bibfnamefont {M.-J.}\ \bibnamefont
  {Zhao}}, \bibinfo {author} {\bibfnamefont {X.-J.}\ \bibnamefont {Bi}},\ and\
  \bibinfo {author} {\bibfnamefont {K.}~\bibnamefont {Fang}},\ }\href
  {https://doi.org/10.1103/PhysRevD.107.063020} {\bibfield  {journal} {\bibinfo
   {journal} {Phys. Rev. D}\ }\textbf {\bibinfo {volume} {107}},\ \bibinfo
  {pages} {063020} (\bibinfo {year} {2023})},\ \Eprint
  {https://arxiv.org/abs/2209.03799} {arxiv:2209.03799 [astro-ph.HE]}
  \BibitemShut {NoStop}%
\bibitem [{\citenamefont {Orlando}\ and\ \citenamefont
  {Strong}(2013)}]{Orlando:2013ysa}%
  \BibitemOpen
  \bibfield  {author} {\bibinfo {author} {\bibfnamefont {E.}~\bibnamefont
  {Orlando}}\ and\ \bibinfo {author} {\bibfnamefont {A.}~\bibnamefont
  {Strong}},\ }\href {https://doi.org/10.1093/mnras/stt1718} {\bibfield
  {journal} {\bibinfo  {journal} {Monthly Notices of the Royal Astronomical
  Society}\ }\textbf {\bibinfo {volume} {436}},\ \bibinfo {pages} {2127}
  (\bibinfo {year} {2013})},\ \Eprint {https://arxiv.org/abs/1309.2947}
  {arxiv:1309.2947 [astro-ph.GA]} \BibitemShut {NoStop}%
\bibitem [{\citenamefont {Tomassetti}(2015)}]{Tomassetti:2015nha}%
  \BibitemOpen
  \bibfield  {author} {\bibinfo {author} {\bibfnamefont {N.}~\bibnamefont
  {Tomassetti}},\ }\href {https://doi.org/10.1103/PhysRevC.92.045808}
  {\bibfield  {journal} {\bibinfo  {journal} {Physical Review C}\ }\textbf
  {\bibinfo {volume} {92}},\ \bibinfo {pages} {045808} (\bibinfo {year}
  {2015})},\ \Eprint {https://arxiv.org/abs/1509.05776} {arxiv:1509.05776
  [astro-ph.HE]} \BibitemShut {NoStop}%
\bibitem [{\citenamefont {J{\'o}hannesson}\ \emph {et~al.}(2016)\citenamefont
  {J{\'o}hannesson}, \citenamefont {{de Austri}}, \citenamefont {Vincent},
  \citenamefont {Moskalenko}, \citenamefont {Orlando}, \citenamefont {Porter},
  \citenamefont {Strong}, \citenamefont {Trotta}, \citenamefont {Feroz},
  \citenamefont {Graff},\ and\ \citenamefont {Hobson}}]{Johannesson:2016rlh}%
  \BibitemOpen
  \bibfield  {author} {\bibinfo {author} {\bibfnamefont {G.}~\bibnamefont
  {J{\'o}hannesson}}, \bibinfo {author} {\bibfnamefont {R.~R.}\ \bibnamefont
  {{de Austri}}}, \bibinfo {author} {\bibfnamefont {A.~C.}\ \bibnamefont
  {Vincent}}, \bibinfo {author} {\bibfnamefont {I.~V.}\ \bibnamefont
  {Moskalenko}}, \bibinfo {author} {\bibfnamefont {E.}~\bibnamefont {Orlando}},
  \bibinfo {author} {\bibfnamefont {T.~A.}\ \bibnamefont {Porter}}, \bibinfo
  {author} {\bibfnamefont {A.~W.}\ \bibnamefont {Strong}}, \bibinfo {author}
  {\bibfnamefont {R.}~\bibnamefont {Trotta}}, \bibinfo {author} {\bibfnamefont
  {F.}~\bibnamefont {Feroz}}, \bibinfo {author} {\bibfnamefont
  {P.}~\bibnamefont {Graff}},\ and\ \bibinfo {author} {\bibfnamefont {M.~P.}\
  \bibnamefont {Hobson}},\ }\href {https://doi.org/10.3847/0004-637X/824/1/16}
  {\bibfield  {journal} {\bibinfo  {journal} {The Astrophysical Journal}\
  }\textbf {\bibinfo {volume} {824}},\ \bibinfo {pages} {16} (\bibinfo {year}
  {2016})},\ \Eprint {https://arxiv.org/abs/1602.02243} {arxiv:1602.02243
  [astro-ph.HE]} \BibitemShut {NoStop}%
\bibitem [{\citenamefont {Winkler}(2017)}]{Winkler:2017xor}%
  \BibitemOpen
  \bibfield  {author} {\bibinfo {author} {\bibfnamefont {M.~W.}\ \bibnamefont
  {Winkler}},\ }\href {https://doi.org/10.1088/1475-7516/2017/02/048}
  {\bibfield  {journal} {\bibinfo  {journal} {Journal of Cosmology and
  Astroparticle Physics}\ }\textbf {\bibinfo {volume} {02}}\bibfield  {number}
  {\bibinfo  {number} { (02)},\ \bibinfo {pages} {048}},\ }\Eprint
  {https://arxiv.org/abs/1701.04866} {arxiv:1701.04866 [hep-ph]} \BibitemShut
  {NoStop}%
\bibitem [{\citenamefont {Aguilar}\ \emph {et~al.}(2021)\citenamefont {Aguilar}
  \emph {et~al.}}]{AMS:2021nhj}%
  \BibitemOpen
  \bibfield  {author} {\bibinfo {author} {\bibfnamefont {M.}~\bibnamefont
  {Aguilar}} \emph {et~al.} (\bibinfo {collaboration} {AMS}),\ }\href
  {https://doi.org/10.1016/j.physrep.2020.09.003} {\bibfield  {journal}
  {\bibinfo  {journal} {Physics Reports}\ }\bibinfo {series} {The {{Alpha
  Magnetic Spectrometer}} ({{AMS}}) on the {{International Space Station}}:
  {{Part II}} - {{Results}} from the {{First Seven Years}}},\ \textbf {\bibinfo
  {volume} {894}},\ \bibinfo {pages} {1} (\bibinfo {year} {2021})}\BibitemShut
  {NoStop}%
\bibitem [{\citenamefont {Cummings}\ \emph {et~al.}(2016)\citenamefont
  {Cummings}, \citenamefont {Stone}, \citenamefont {Heikkila}, \citenamefont
  {Lal}, \citenamefont {Webber}, \citenamefont {J{\'o}hannesson}, \citenamefont
  {Moskalenko}, \citenamefont {Orlando},\ and\ \citenamefont
  {Porter}}]{Cummings:2016pdr}%
  \BibitemOpen
  \bibfield  {author} {\bibinfo {author} {\bibfnamefont {A.~C.}\ \bibnamefont
  {Cummings}}, \bibinfo {author} {\bibfnamefont {E.~C.}\ \bibnamefont {Stone}},
  \bibinfo {author} {\bibfnamefont {B.~C.}\ \bibnamefont {Heikkila}}, \bibinfo
  {author} {\bibfnamefont {N.}~\bibnamefont {Lal}}, \bibinfo {author}
  {\bibfnamefont {W.~R.}\ \bibnamefont {Webber}}, \bibinfo {author}
  {\bibfnamefont {G.}~\bibnamefont {J{\'o}hannesson}}, \bibinfo {author}
  {\bibfnamefont {I.~V.}\ \bibnamefont {Moskalenko}}, \bibinfo {author}
  {\bibfnamefont {E.}~\bibnamefont {Orlando}},\ and\ \bibinfo {author}
  {\bibfnamefont {T.~A.}\ \bibnamefont {Porter}},\ }\href
  {https://doi.org/10.3847/0004-637X/831/1/18} {\bibfield  {journal} {\bibinfo
  {journal} {The Astrophysical Journal}\ }\textbf {\bibinfo {volume} {831}},\
  \bibinfo {pages} {18} (\bibinfo {year} {2016})}\BibitemShut {NoStop}%
\bibitem [{\citenamefont {An}\ \emph {et~al.}(2019)\citenamefont {An} \emph
  {et~al.}}]{DAMPE:2019gys}%
  \BibitemOpen
  \bibfield  {author} {\bibinfo {author} {\bibfnamefont {Q.}~\bibnamefont {An}}
  \emph {et~al.} (\bibinfo {collaboration} {DAMPE}),\ }\href
  {https://doi.org/10.1126/sciadv.aax3793} {\bibfield  {journal} {\bibinfo
  {journal} {Science Advances}\ }\textbf {\bibinfo {volume} {5}},\ \bibinfo
  {pages} {eaax3793} (\bibinfo {year} {2019})},\ \Eprint
  {https://arxiv.org/abs/1909.12860} {arxiv:1909.12860 [astro-ph.HE]}
  \BibitemShut {NoStop}%
\bibitem [{\citenamefont {Lewis}(2013)}]{Lewis:2013hha}%
  \BibitemOpen
  \bibfield  {author} {\bibinfo {author} {\bibfnamefont {A.}~\bibnamefont
  {Lewis}},\ }\href {https://doi.org/10.1103/PhysRevD.87.103529} {\bibfield
  {journal} {\bibinfo  {journal} {Phys. Rev.}\ }\textbf {\bibinfo {volume}
  {D87}},\ \bibinfo {pages} {103529} (\bibinfo {year} {2013})},\ \Eprint
  {https://arxiv.org/abs/1304.4473} {arXiv:1304.4473 [astro-ph.CO]}
  \BibitemShut {NoStop}%
\bibitem [{\citenamefont {Torrado}\ and\ \citenamefont
  {Lewis}(2021)}]{Torrado:2020dgo}%
  \BibitemOpen
  \bibfield  {author} {\bibinfo {author} {\bibfnamefont {J.}~\bibnamefont
  {Torrado}}\ and\ \bibinfo {author} {\bibfnamefont {A.}~\bibnamefont
  {Lewis}},\ }\href {https://doi.org/10.1088/1475-7516/2021/05/057} {\bibfield
  {journal} {\bibinfo  {journal} {Journal of Cosmology and Astroparticle
  Physics}\ }\textbf {\bibinfo {volume} {2021}}\bibfield  {number} {\bibinfo
  {number} { (05)},\ \bibinfo {pages} {057}},\ }\Eprint
  {https://arxiv.org/abs/2005.05290} {arxiv:2005.05290 [astro-ph.IM]}
  \BibitemShut {NoStop}%
\bibitem [{\citenamefont {{Lewis}}(2019)}]{Lewis:2019xzd}%
  \BibitemOpen
  \bibfield  {author} {\bibinfo {author} {\bibfnamefont {A.}~\bibnamefont
  {{Lewis}}},\ }\href@noop {} {\bibfield  {journal} {\bibinfo  {journal} {arXiv
  e-prints}\ ,\ \bibinfo {eid} {arXiv:1910.13970}} (\bibinfo {year} {2019})},\
  \Eprint {https://arxiv.org/abs/1910.13970} {arXiv:1910.13970 [astro-ph.IM]}
  \BibitemShut {NoStop}%
\bibitem [{\citenamefont {Tan}\ and\ \citenamefont {Ng}(1983)}]{Tan:1983kgh}%
  \BibitemOpen
  \bibfield  {author} {\bibinfo {author} {\bibfnamefont {L.~C.}\ \bibnamefont
  {Tan}}\ and\ \bibinfo {author} {\bibfnamefont {L.~K.}\ \bibnamefont {Ng}},\
  }\href {https://doi.org/10.1088/0305-4616/9/10/015} {\bibfield  {journal}
  {\bibinfo  {journal} {Journal of Physics G: Nuclear Physics}\ }\textbf
  {\bibinfo {volume} {9}},\ \bibinfo {pages} {1289} (\bibinfo {year}
  {1983})}\BibitemShut {NoStop}%
\bibitem [{\citenamefont {Kachelriess}\ \emph {et~al.}(2015)\citenamefont
  {Kachelriess}, \citenamefont {Moskalenko},\ and\ \citenamefont
  {Ostapchenko}}]{Kachelriess:2015wpa}%
  \BibitemOpen
  \bibfield  {author} {\bibinfo {author} {\bibfnamefont {M.}~\bibnamefont
  {Kachelriess}}, \bibinfo {author} {\bibfnamefont {I.~V.}\ \bibnamefont
  {Moskalenko}},\ and\ \bibinfo {author} {\bibfnamefont {S.~S.}\ \bibnamefont
  {Ostapchenko}},\ }\href {https://doi.org/10.1088/0004-637X/803/2/54}
  {\bibfield  {journal} {\bibinfo  {journal} {The Astrophysical Journal}\
  }\textbf {\bibinfo {volume} {803}},\ \bibinfo {pages} {54} (\bibinfo {year}
  {2015})},\ \Eprint {https://arxiv.org/abs/1502.04158} {arxiv:1502.04158
  [astro-ph.HE]} \BibitemShut {NoStop}%
\bibitem [{\citenamefont {Pierog}\ and\ \citenamefont
  {Werner}(2009)}]{Pierog:2009zt}%
  \BibitemOpen
  \bibfield  {author} {\bibinfo {author} {\bibfnamefont {T.}~\bibnamefont
  {Pierog}}\ and\ \bibinfo {author} {\bibfnamefont {K.}~\bibnamefont
  {Werner}},\ }\href {https://doi.org/10.1016/j.nuclphysbps.2009.09.017}
  {\bibfield  {journal} {\bibinfo  {journal} {Nuclear Physics B - Proceedings
  Supplements}\ }\textbf {\bibinfo {volume} {196}},\ \bibinfo {pages} {102}
  (\bibinfo {year} {2009})},\ \Eprint {https://arxiv.org/abs/0905.1198}
  {arxiv:0905.1198 [hep-ph]} \BibitemShut {NoStop}%
\bibitem [{\citenamefont {Aguilar}\ \emph {et~al.}(2016)\citenamefont {Aguilar}
  \emph {et~al.}}]{AMS:2016oqu}%
  \BibitemOpen
  \bibfield  {author} {\bibinfo {author} {\bibfnamefont {M.}~\bibnamefont
  {Aguilar}} \emph {et~al.} (\bibinfo {collaboration} {AMS}),\ }\href
  {https://doi.org/10.1103/PhysRevLett.117.091103} {\bibfield  {journal}
  {\bibinfo  {journal} {Physical Review Letters}\ }\textbf {\bibinfo {volume}
  {117}},\ \bibinfo {pages} {091103} (\bibinfo {year} {2016})}\BibitemShut
  {NoStop}%
\bibitem [{\citenamefont {Cirelli}\ \emph {et~al.}(2011)\citenamefont
  {Cirelli}, \citenamefont {Corcella}, \citenamefont {Hektor}, \citenamefont
  {H{\"u}tsi}, \citenamefont {Kadastik}, \citenamefont {Panci}, \citenamefont
  {Raidal}, \citenamefont {Sala},\ and\ \citenamefont
  {Strumia}}]{Cirelli:2010xx}%
  \BibitemOpen
  \bibfield  {author} {\bibinfo {author} {\bibfnamefont {M.}~\bibnamefont
  {Cirelli}}, \bibinfo {author} {\bibfnamefont {G.}~\bibnamefont {Corcella}},
  \bibinfo {author} {\bibfnamefont {A.}~\bibnamefont {Hektor}}, \bibinfo
  {author} {\bibfnamefont {G.}~\bibnamefont {H{\"u}tsi}}, \bibinfo {author}
  {\bibfnamefont {M.}~\bibnamefont {Kadastik}}, \bibinfo {author}
  {\bibfnamefont {P.}~\bibnamefont {Panci}}, \bibinfo {author} {\bibfnamefont
  {M.}~\bibnamefont {Raidal}}, \bibinfo {author} {\bibfnamefont
  {F.}~\bibnamefont {Sala}},\ and\ \bibinfo {author} {\bibfnamefont
  {A.}~\bibnamefont {Strumia}},\ }\href
  {https://doi.org/10.1088/1475-7516/2012/10/E01} {\bibfield  {journal}
  {\bibinfo  {journal} {Journal of Cosmology and Astroparticle Physics}\
  }\textbf {\bibinfo {volume} {03}}\bibfield  {number} {\bibinfo  {number} {
  (03)},\ \bibinfo {pages} {051}},\ }\Eprint {https://arxiv.org/abs/1012.4515}
  {arxiv:1012.4515 [hep-ph]} \BibitemShut {NoStop}%
\bibitem [{\citenamefont {Navarro}\ \emph {et~al.}(1997)\citenamefont
  {Navarro}, \citenamefont {Frenk},\ and\ \citenamefont
  {White}}]{Navarro:1996gj}%
  \BibitemOpen
  \bibfield  {author} {\bibinfo {author} {\bibfnamefont {J.~F.}\ \bibnamefont
  {Navarro}}, \bibinfo {author} {\bibfnamefont {C.~S.}\ \bibnamefont {Frenk}},\
  and\ \bibinfo {author} {\bibfnamefont {S.~D.~M.}\ \bibnamefont {White}},\
  }\href {https://doi.org/10.1086/304888} {\bibfield  {journal} {\bibinfo
  {journal} {The Astrophysical Journal}\ }\textbf {\bibinfo {volume} {490}},\
  \bibinfo {pages} {493} (\bibinfo {year} {1997})},\ \Eprint
  {https://arxiv.org/abs/astro-ph/9611107} {arxiv:astro-ph/9611107}
  \BibitemShut {NoStop}%
\bibitem [{\citenamefont {Karukes}\ \emph {et~al.}(2019)\citenamefont
  {Karukes}, \citenamefont {Benito}, \citenamefont {Iocco}, \citenamefont
  {Trotta},\ and\ \citenamefont {{Geringer-Sameth}}}]{Karukes:2019jxv}%
  \BibitemOpen
  \bibfield  {author} {\bibinfo {author} {\bibfnamefont {E.~V.}\ \bibnamefont
  {Karukes}}, \bibinfo {author} {\bibfnamefont {M.}~\bibnamefont {Benito}},
  \bibinfo {author} {\bibfnamefont {F.}~\bibnamefont {Iocco}}, \bibinfo
  {author} {\bibfnamefont {R.}~\bibnamefont {Trotta}},\ and\ \bibinfo {author}
  {\bibfnamefont {A.}~\bibnamefont {{Geringer-Sameth}}},\ }\href
  {https://doi.org/10.1088/1475-7516/2019/09/046} {\bibfield  {journal}
  {\bibinfo  {journal} {Journal of Cosmology and Astroparticle Physics}\
  }\textbf {\bibinfo {volume} {09}}\bibfield  {number} {\bibinfo  {number} {
  (09)},\ \bibinfo {pages} {046}},\ }\Eprint {https://arxiv.org/abs/1901.02463}
  {arxiv:1901.02463 [astro-ph.GA]} \BibitemShut {NoStop}%
\bibitem [{\citenamefont {Benito}\ \emph {et~al.}(2019)\citenamefont {Benito},
  \citenamefont {Cuoco},\ and\ \citenamefont {Iocco}}]{Benito:2019ngh}%
  \BibitemOpen
  \bibfield  {author} {\bibinfo {author} {\bibfnamefont {M.}~\bibnamefont
  {Benito}}, \bibinfo {author} {\bibfnamefont {A.}~\bibnamefont {Cuoco}},\ and\
  \bibinfo {author} {\bibfnamefont {F.}~\bibnamefont {Iocco}},\ }\href
  {https://doi.org/10.1088/1475-7516/2019/03/033} {\bibfield  {journal}
  {\bibinfo  {journal} {Journal of Cosmology and Astroparticle Physics}\
  }\textbf {\bibinfo {volume} {03}}\bibfield  {number} {\bibinfo  {number} {
  (03)},\ \bibinfo {pages} {033}},\ }\Eprint {https://arxiv.org/abs/1901.02460}
  {arxiv:1901.02460 [astro-ph.GA]} \BibitemShut {NoStop}%
\bibitem [{\citenamefont {Merritt}\ \emph {et~al.}(2006)\citenamefont
  {Merritt}, \citenamefont {Graham}, \citenamefont {Moore}, \citenamefont
  {Diemand},\ and\ \citenamefont {Terzic}}]{Graham:2005xx}%
  \BibitemOpen
  \bibfield  {author} {\bibinfo {author} {\bibfnamefont {D.}~\bibnamefont
  {Merritt}}, \bibinfo {author} {\bibfnamefont {A.~W.}\ \bibnamefont {Graham}},
  \bibinfo {author} {\bibfnamefont {B.}~\bibnamefont {Moore}}, \bibinfo
  {author} {\bibfnamefont {J.}~\bibnamefont {Diemand}},\ and\ \bibinfo {author}
  {\bibfnamefont {B.}~\bibnamefont {Terzic}},\ }\href
  {https://doi.org/10.1086/508988} {\bibfield  {journal} {\bibinfo  {journal}
  {The Astronomical Journal}\ }\textbf {\bibinfo {volume} {132}},\ \bibinfo
  {pages} {2685} (\bibinfo {year} {2006})},\ \Eprint
  {https://arxiv.org/abs/astro-ph/0509417} {arxiv:astro-ph/0509417}
  \BibitemShut {NoStop}%
\bibitem [{\citenamefont {Navarro}\ \emph {et~al.}(2010)\citenamefont
  {Navarro}, \citenamefont {Ludlow}, \citenamefont {Springel}, \citenamefont
  {Wang}, \citenamefont {Vogelsberger}, \citenamefont {White}, \citenamefont
  {Jenkins}, \citenamefont {Frenk},\ and\ \citenamefont
  {Helmi}}]{Navarro:2008kc}%
  \BibitemOpen
  \bibfield  {author} {\bibinfo {author} {\bibfnamefont {J.~F.}\ \bibnamefont
  {Navarro}}, \bibinfo {author} {\bibfnamefont {A.}~\bibnamefont {Ludlow}},
  \bibinfo {author} {\bibfnamefont {V.}~\bibnamefont {Springel}}, \bibinfo
  {author} {\bibfnamefont {J.}~\bibnamefont {Wang}}, \bibinfo {author}
  {\bibfnamefont {M.}~\bibnamefont {Vogelsberger}}, \bibinfo {author}
  {\bibfnamefont {S.~D.~M.}\ \bibnamefont {White}}, \bibinfo {author}
  {\bibfnamefont {A.}~\bibnamefont {Jenkins}}, \bibinfo {author} {\bibfnamefont
  {C.~S.}\ \bibnamefont {Frenk}},\ and\ \bibinfo {author} {\bibfnamefont
  {A.}~\bibnamefont {Helmi}},\ }\href
  {https://doi.org/10.1111/j.1365-2966.2009.15878.x} {\bibfield  {journal}
  {\bibinfo  {journal} {Monthly Notices of the Royal Astronomical Society}\
  }\textbf {\bibinfo {volume} {402}},\ \bibinfo {pages} {21} (\bibinfo {year}
  {2010})},\ \Eprint {https://arxiv.org/abs/0810.1522} {arxiv:0810.1522
  [astro-ph]} \BibitemShut {NoStop}%
\bibitem [{\citenamefont {Burkert}(1996)}]{Burkert:1995yz}%
  \BibitemOpen
  \bibfield  {author} {\bibinfo {author} {\bibfnamefont {A.}~\bibnamefont
  {Burkert}},\ }\href {https://doi.org/10.1086/309560} {\bibfield  {journal}
  {\bibinfo  {journal} {The Astrophysical Journal}\ }\textbf {\bibinfo {volume}
  {171}},\ \bibinfo {pages} {175} (\bibinfo {year} {1996})},\ \Eprint
  {https://arxiv.org/abs/astro-ph/9504041} {arxiv:astro-ph/9504041}
  \BibitemShut {NoStop}%
\bibitem [{\citenamefont {Salucci}\ and\ \citenamefont
  {Burkert}(2000)}]{Salucci:2000ps}%
  \BibitemOpen
  \bibfield  {author} {\bibinfo {author} {\bibfnamefont {P.}~\bibnamefont
  {Salucci}}\ and\ \bibinfo {author} {\bibfnamefont {A.}~\bibnamefont
  {Burkert}},\ }\href {https://doi.org/10.1086/312747} {\bibfield  {journal}
  {\bibinfo  {journal} {The Astrophysical Journal}\ }\textbf {\bibinfo {volume}
  {537}},\ \bibinfo {pages} {L9} (\bibinfo {year} {2000})},\ \Eprint
  {https://arxiv.org/abs/astro-ph/0004397} {arxiv:astro-ph/0004397}
  \BibitemShut {NoStop}%
\bibitem [{\citenamefont {Cuoco}\ \emph {et~al.}(2018)\citenamefont {Cuoco},
  \citenamefont {Heisig}, \citenamefont {Korsmeier},\ and\ \citenamefont
  {Kr{\"a}mer}}]{Cuoco:2017iax}%
  \BibitemOpen
  \bibfield  {author} {\bibinfo {author} {\bibfnamefont {A.}~\bibnamefont
  {Cuoco}}, \bibinfo {author} {\bibfnamefont {J.}~\bibnamefont {Heisig}},
  \bibinfo {author} {\bibfnamefont {M.}~\bibnamefont {Korsmeier}},\ and\
  \bibinfo {author} {\bibfnamefont {M.}~\bibnamefont {Kr{\"a}mer}},\ }\href
  {https://doi.org/10.1088/1475-7516/2018/04/004} {\bibfield  {journal}
  {\bibinfo  {journal} {JCAP}\ }\textbf {\bibinfo {volume} {2018}}\bibfield
  {number} {\bibinfo  {number} { (04)},\ \bibinfo {pages} {004}},\ }\Eprint
  {https://arxiv.org/abs/1711.05274} {arxiv:1711.05274 [hep-ph]} \BibitemShut
  {NoStop}%
\bibitem [{\citenamefont {Derome}\ \emph {et~al.}(2019)\citenamefont {Derome},
  \citenamefont {Maurin}, \citenamefont {Salati}, \citenamefont {Boudaud},
  \citenamefont {G{\'e}nolini},\ and\ \citenamefont
  {Kunz{\'e}}}]{Derome:2019jfs}%
  \BibitemOpen
  \bibfield  {author} {\bibinfo {author} {\bibfnamefont {L.}~\bibnamefont
  {Derome}}, \bibinfo {author} {\bibfnamefont {D.}~\bibnamefont {Maurin}},
  \bibinfo {author} {\bibfnamefont {P.}~\bibnamefont {Salati}}, \bibinfo
  {author} {\bibfnamefont {M.}~\bibnamefont {Boudaud}}, \bibinfo {author}
  {\bibfnamefont {Y.}~\bibnamefont {G{\'e}nolini}},\ and\ \bibinfo {author}
  {\bibfnamefont {P.}~\bibnamefont {Kunz{\'e}}},\ }\href
  {https://doi.org/10.1051/0004-6361/201935717} {\bibfield  {journal} {\bibinfo
   {journal} {Astronomy \& Astrophysics}\ }\textbf {\bibinfo {volume} {627}},\
  \bibinfo {pages} {A158} (\bibinfo {year} {2019})},\ \Eprint
  {https://arxiv.org/abs/1904.08210} {arxiv:1904.08210 [astro-ph.HE]}
  \BibitemShut {NoStop}%
\bibitem [{\citenamefont {Jin}\ \emph {et~al.}(2015)\citenamefont {Jin},
  \citenamefont {Wu},\ and\ \citenamefont {Zhou}}]{Jin:2014ica}%
  \BibitemOpen
  \bibfield  {author} {\bibinfo {author} {\bibfnamefont {H.-B.}\ \bibnamefont
  {Jin}}, \bibinfo {author} {\bibfnamefont {Y.-L.}\ \bibnamefont {Wu}},\ and\
  \bibinfo {author} {\bibfnamefont {Y.-F.}\ \bibnamefont {Zhou}},\ }\href
  {https://doi.org/10.1088/1475-7516/2015/09/049} {\bibfield  {journal}
  {\bibinfo  {journal} {Journal of Cosmology and Astroparticle Physics}\
  }\textbf {\bibinfo {volume} {09}}\bibfield  {number} {\bibinfo  {number} {
  (09)},\ \bibinfo {pages} {049}},\ }\Eprint {https://arxiv.org/abs/1410.0171}
  {arxiv:1410.0171 [hep-ph]} \BibitemShut {NoStop}%
\bibitem [{\citenamefont {Brooks}\ \emph {et~al.}(2011)\citenamefont {Brooks},
  \citenamefont {Gelman}, \citenamefont {Jones},\ and\ \citenamefont
  {Meng}}]{brooksHandbookMarkovChain2011}%
  \BibitemOpen
  \bibinfo {editor} {\bibfnamefont {S.}~\bibnamefont {Brooks}}, \bibinfo
  {editor} {\bibfnamefont {A.}~\bibnamefont {Gelman}}, \bibinfo {editor}
  {\bibfnamefont {G.}~\bibnamefont {Jones}},\ and\ \bibinfo {editor}
  {\bibfnamefont {X.-L.}\ \bibnamefont {Meng}},\ eds.,\ \href
  {https://doi.org/10.1201/b10905} {\emph {\bibinfo {title} {Handbook of
  {{Markov Chain Monte Carlo}}}}}\ (\bibinfo  {publisher} {{Chapman and
  Hall/CRC}},\ \bibinfo {address} {{New York}},\ \bibinfo {year}
  {2011})\BibitemShut {NoStop}%
\bibitem [{\citenamefont {Collaborations}(2017)}]{Fermi-LAT:2016uux}%
  \BibitemOpen
  \bibfield  {author} {\bibinfo {author} {\bibfnamefont {F.-L.}\ \bibnamefont
  {Collaborations}} (\bibinfo {collaboration} {Fermi-LAT, DES}),\ }\href
  {https://doi.org/10.3847/1538-4357/834/2/110} {\bibfield  {journal} {\bibinfo
   {journal} {The Astrophysical Journal}\ }\textbf {\bibinfo {volume} {834}},\
  \bibinfo {pages} {110} (\bibinfo {year} {2017})},\ \Eprint
  {https://arxiv.org/abs/1611.03184} {arxiv:1611.03184 [astro-ph.HE]}
  \BibitemShut {NoStop}%
\bibitem [{\citenamefont {{de Salas}}\ and\ \citenamefont
  {Widmark}(2021)}]{deSalas:2020hbh}%
  \BibitemOpen
  \bibfield  {author} {\bibinfo {author} {\bibfnamefont {P.~F.}\ \bibnamefont
  {{de Salas}}}\ and\ \bibinfo {author} {\bibfnamefont {A.}~\bibnamefont
  {Widmark}},\ }\href {https://doi.org/10.1088/1361-6633/ac24e7} {\bibfield
  {journal} {\bibinfo  {journal} {Rept. Prog. Phys.}\ }\textbf {\bibinfo
  {volume} {84}},\ \bibinfo {pages} {104901} (\bibinfo {year} {2021})},\
  \Eprint {https://arxiv.org/abs/2012.11477} {arxiv:2012.11477 [astro-ph.GA]}
  \BibitemShut {NoStop}%
\bibitem [{\citenamefont {Gross}\ and\ \citenamefont
  {Vitells}(2010)}]{Gross:2010qma}%
  \BibitemOpen
  \bibfield  {author} {\bibinfo {author} {\bibfnamefont {E.}~\bibnamefont
  {Gross}}\ and\ \bibinfo {author} {\bibfnamefont {O.}~\bibnamefont
  {Vitells}},\ }\href {https://doi.org/10.1140/epjc/s10052-010-1470-8}
  {\bibfield  {journal} {\bibinfo  {journal} {The European Physical Journal C}\
  }\textbf {\bibinfo {volume} {70}},\ \bibinfo {pages} {525} (\bibinfo {year}
  {2010})},\ \Eprint {https://arxiv.org/abs/1005.1891} {arxiv:1005.1891
  [physics.data-an]} \BibitemShut {NoStop}%
\bibitem [{\citenamefont {Steigman}\ \emph {et~al.}(2012)\citenamefont
  {Steigman}, \citenamefont {Dasgupta},\ and\ \citenamefont
  {Beacom}}]{Steigman:2012nb}%
  \BibitemOpen
  \bibfield  {author} {\bibinfo {author} {\bibfnamefont {G.}~\bibnamefont
  {Steigman}}, \bibinfo {author} {\bibfnamefont {B.}~\bibnamefont {Dasgupta}},\
  and\ \bibinfo {author} {\bibfnamefont {J.~F.}\ \bibnamefont {Beacom}},\
  }\href {https://doi.org/10.1103/PhysRevD.86.023506} {\bibfield  {journal}
  {\bibinfo  {journal} {Physical Review D}\ }\textbf {\bibinfo {volume} {86}},\
  \bibinfo {pages} {023506} (\bibinfo {year} {2012})},\ \Eprint
  {https://arxiv.org/abs/1204.3622} {arxiv:1204.3622 [hep-ph]} \BibitemShut
  {NoStop}%
\bibitem [{\citenamefont {Heisig}(2020)}]{Heisig:2020jvs}%
  \BibitemOpen
  \bibfield  {author} {\bibinfo {author} {\bibfnamefont {J.}~\bibnamefont
  {Heisig}},\ }\href {https://doi.org/10.1142/S0217732321300032} {\bibfield
  {journal} {\bibinfo  {journal} {Mod. Phys. Lett. A}\ }\textbf {\bibinfo
  {volume} {36}},\ \bibinfo {pages} {2130003} (\bibinfo {year} {2020})},\
  \Eprint {https://arxiv.org/abs/2012.03956} {arxiv:2012.03956 [astro-ph.HE]}
  \BibitemShut {NoStop}%
\bibitem [{\citenamefont {Hooper}\ and\ \citenamefont
  {Slatyer}(2013)}]{Hooper:2013rwa}%
  \BibitemOpen
  \bibfield  {author} {\bibinfo {author} {\bibfnamefont {D.}~\bibnamefont
  {Hooper}}\ and\ \bibinfo {author} {\bibfnamefont {T.~R.}\ \bibnamefont
  {Slatyer}},\ }\href {https://doi.org/10.1016/j.dark.2013.06.003} {\bibfield
  {journal} {\bibinfo  {journal} {Physics of the Dark Universe}\ }\textbf
  {\bibinfo {volume} {2}},\ \bibinfo {pages} {118} (\bibinfo {year} {2013})},\
  \Eprint {https://arxiv.org/abs/1302.6589} {arxiv:1302.6589 [astro-ph.HE]}
  \BibitemShut {NoStop}%
\bibitem [{\citenamefont {Gordon}\ and\ \citenamefont
  {Macias}(2013)}]{Gordon:2013vta}%
  \BibitemOpen
  \bibfield  {author} {\bibinfo {author} {\bibfnamefont {C.}~\bibnamefont
  {Gordon}}\ and\ \bibinfo {author} {\bibfnamefont {O.}~\bibnamefont
  {Macias}},\ }\href {https://doi.org/10.1103/PhysRevD.88.083521} {\bibfield
  {journal} {\bibinfo  {journal} {Physical Review D}\ }\textbf {\bibinfo
  {volume} {88}},\ \bibinfo {pages} {083521} (\bibinfo {year} {2013})},\
  \Eprint {https://arxiv.org/abs/1306.5725} {arxiv:1306.5725 [astro-ph.HE]}
  \BibitemShut {NoStop}%
\bibitem [{\citenamefont {{Huang}}\ \emph {et~al.}(2013)\citenamefont
  {{Huang}}, \citenamefont {{Urbano}},\ and\ \citenamefont
  {{Xue}}}]{2013arXiv1307.6862H}%
  \BibitemOpen
  \bibfield  {author} {\bibinfo {author} {\bibfnamefont {W.-C.}\ \bibnamefont
  {{Huang}}}, \bibinfo {author} {\bibfnamefont {A.}~\bibnamefont {{Urbano}}},\
  and\ \bibinfo {author} {\bibfnamefont {W.}~\bibnamefont {{Xue}}},\ }\href
  {https://doi.org/10.48550/arXiv.1307.6862} {\bibfield  {journal} {\bibinfo
  {journal} {arXiv e-prints}\ ,\ \bibinfo {eid} {arXiv:1307.6862}} (\bibinfo
  {year} {2013})},\ \Eprint {https://arxiv.org/abs/1307.6862} {arXiv:1307.6862
  [hep-ph]} \BibitemShut {NoStop}%
\bibitem [{\citenamefont {Daylan}\ \emph {et~al.}(2016)\citenamefont {Daylan},
  \citenamefont {Finkbeiner}, \citenamefont {Hooper}, \citenamefont {Linden},
  \citenamefont {Portillo}, \citenamefont {Rodd},\ and\ \citenamefont
  {Slatyer}}]{Daylan:2014rsa}%
  \BibitemOpen
  \bibfield  {author} {\bibinfo {author} {\bibfnamefont {T.}~\bibnamefont
  {Daylan}}, \bibinfo {author} {\bibfnamefont {D.~P.}\ \bibnamefont
  {Finkbeiner}}, \bibinfo {author} {\bibfnamefont {D.}~\bibnamefont {Hooper}},
  \bibinfo {author} {\bibfnamefont {T.}~\bibnamefont {Linden}}, \bibinfo
  {author} {\bibfnamefont {S.~K.~N.}\ \bibnamefont {Portillo}}, \bibinfo
  {author} {\bibfnamefont {N.~L.}\ \bibnamefont {Rodd}},\ and\ \bibinfo
  {author} {\bibfnamefont {T.~R.}\ \bibnamefont {Slatyer}},\ }\href
  {https://doi.org/10.1016/j.dark.2015.12.005} {\bibfield  {journal} {\bibinfo
  {journal} {Physics of the Dark Universe}\ }\textbf {\bibinfo {volume} {12}},\
  \bibinfo {pages} {1} (\bibinfo {year} {2016})},\ \Eprint
  {https://arxiv.org/abs/1402.6703} {arxiv:1402.6703 [astro-ph.HE]}
  \BibitemShut {NoStop}%
\bibitem [{\citenamefont {Abazajian}\ \emph {et~al.}(2014)\citenamefont
  {Abazajian}, \citenamefont {Canac}, \citenamefont {Horiuchi},\ and\
  \citenamefont {Kaplinghat}}]{Abazajian:2014fta}%
  \BibitemOpen
  \bibfield  {author} {\bibinfo {author} {\bibfnamefont {K.~N.}\ \bibnamefont
  {Abazajian}}, \bibinfo {author} {\bibfnamefont {N.}~\bibnamefont {Canac}},
  \bibinfo {author} {\bibfnamefont {S.}~\bibnamefont {Horiuchi}},\ and\
  \bibinfo {author} {\bibfnamefont {M.}~\bibnamefont {Kaplinghat}},\ }\href
  {https://doi.org/10.1103/PhysRevD.90.023526} {\bibfield  {journal} {\bibinfo
  {journal} {Physical Review D}\ }\textbf {\bibinfo {volume} {90}},\ \bibinfo
  {pages} {023526} (\bibinfo {year} {2014})},\ \Eprint
  {https://arxiv.org/abs/1402.4090} {arxiv:1402.4090 [astro-ph.HE]}
  \BibitemShut {NoStop}%
\bibitem [{\citenamefont {Agrawal}\ \emph {et~al.}(2015)\citenamefont
  {Agrawal}, \citenamefont {Batell}, \citenamefont {Fox},\ and\ \citenamefont
  {Harnik}}]{Agrawal:2014oha}%
  \BibitemOpen
  \bibfield  {author} {\bibinfo {author} {\bibfnamefont {P.}~\bibnamefont
  {Agrawal}}, \bibinfo {author} {\bibfnamefont {B.}~\bibnamefont {Batell}},
  \bibinfo {author} {\bibfnamefont {P.~J.}\ \bibnamefont {Fox}},\ and\ \bibinfo
  {author} {\bibfnamefont {R.}~\bibnamefont {Harnik}},\ }\href
  {https://doi.org/10.1088/1475-7516/2015/05/011} {\bibfield  {journal}
  {\bibinfo  {journal} {Journal of Cosmology and Astroparticle Physics}\
  }\textbf {\bibinfo {volume} {05}}\bibfield  {number} {\bibinfo  {number} {
  (05)},\ \bibinfo {pages} {011}},\ }\Eprint {https://arxiv.org/abs/1411.2592}
  {arxiv:1411.2592 [hep-ph]} \BibitemShut {NoStop}%
\bibitem [{\citenamefont {Calore}\ \emph {et~al.}(2015)\citenamefont {Calore},
  \citenamefont {Cholis},\ and\ \citenamefont {Weniger}}]{Calore:2014xka}%
  \BibitemOpen
  \bibfield  {author} {\bibinfo {author} {\bibfnamefont {F.}~\bibnamefont
  {Calore}}, \bibinfo {author} {\bibfnamefont {I.}~\bibnamefont {Cholis}},\
  and\ \bibinfo {author} {\bibfnamefont {C.}~\bibnamefont {Weniger}},\ }\href
  {https://doi.org/10.1088/1475-7516/2015/03/038} {\bibfield  {journal}
  {\bibinfo  {journal} {Journal of Cosmology and Astroparticle Physics}\
  }\textbf {\bibinfo {volume} {03}}\bibfield  {number} {\bibinfo  {number} {
  (03)},\ \bibinfo {pages} {038}},\ }\Eprint {https://arxiv.org/abs/1409.0042}
  {arxiv:1409.0042 [astro-ph.CO]} \BibitemShut {NoStop}%
\bibitem [{\citenamefont {Cholis}\ \emph {et~al.}(2022)\citenamefont {Cholis},
  \citenamefont {Zhong}, \citenamefont {McDermott},\ and\ \citenamefont
  {Surdutovich}}]{Cholis:2021rpp}%
  \BibitemOpen
  \bibfield  {author} {\bibinfo {author} {\bibfnamefont {I.}~\bibnamefont
  {Cholis}}, \bibinfo {author} {\bibfnamefont {Y.-M.}\ \bibnamefont {Zhong}},
  \bibinfo {author} {\bibfnamefont {S.~D.}\ \bibnamefont {McDermott}},\ and\
  \bibinfo {author} {\bibfnamefont {J.~P.}\ \bibnamefont {Surdutovich}},\
  }\href {https://doi.org/10.1103/PhysRevD.105.103023} {\bibfield  {journal}
  {\bibinfo  {journal} {Physical Review D}\ }\textbf {\bibinfo {volume}
  {105}},\ \bibinfo {pages} {103023} (\bibinfo {year} {2022})},\ \Eprint
  {https://arxiv.org/abs/2112.09706} {arxiv:2112.09706 [astro-ph.HE]}
  \BibitemShut {NoStop}%
\bibitem [{\citenamefont {Zhao}\ \emph {et~al.}(2021)\citenamefont {Zhao},
  \citenamefont {Fang},\ and\ \citenamefont {Bi}}]{Zhao:2021yzf}%
  \BibitemOpen
  \bibfield  {author} {\bibinfo {author} {\bibfnamefont {M.-J.}\ \bibnamefont
  {Zhao}}, \bibinfo {author} {\bibfnamefont {K.}~\bibnamefont {Fang}},\ and\
  \bibinfo {author} {\bibfnamefont {X.-J.}\ \bibnamefont {Bi}},\ }\href
  {https://doi.org/10.1103/PhysRevD.104.123001} {\bibfield  {journal} {\bibinfo
   {journal} {Physical Review D}\ }\textbf {\bibinfo {volume} {104}},\ \bibinfo
  {pages} {123001} (\bibinfo {year} {2021})},\ \Eprint
  {https://arxiv.org/abs/2109.04112} {arxiv:2109.04112 [astro-ph.HE]}
  \BibitemShut {NoStop}%
\bibitem [{\citenamefont {Boschini}\ \emph {et~al.}(2018)\citenamefont
  {Boschini}, \citenamefont {Della~Torre}, \citenamefont {Gervasi},
  \citenamefont {La~Vacca},\ and\ \citenamefont {Rancoita}}]{Boschini:2017gic}%
  \BibitemOpen
  \bibfield  {author} {\bibinfo {author} {\bibfnamefont {M.}~\bibnamefont
  {Boschini}}, \bibinfo {author} {\bibfnamefont {S.}~\bibnamefont
  {Della~Torre}}, \bibinfo {author} {\bibfnamefont {M.}~\bibnamefont
  {Gervasi}}, \bibinfo {author} {\bibfnamefont {G.}~\bibnamefont {La~Vacca}},\
  and\ \bibinfo {author} {\bibfnamefont {P.}~\bibnamefont {Rancoita}},\ }\href
  {https://doi.org/10.1016/j.asr.2017.04.017} {\bibfield  {journal} {\bibinfo
  {journal} {Advances in Space Research}\ }\textbf {\bibinfo {volume} {62}},\
  \bibinfo {pages} {2859} (\bibinfo {year} {2018})},\ \Eprint
  {https://arxiv.org/abs/1704.03733} {arxiv:1704.03733 [astro-ph.SR]}
  \BibitemShut {NoStop}%
\bibitem [{\citenamefont {Kappl}(2016)}]{Kappl:2015hxv}%
  \BibitemOpen
  \bibfield  {author} {\bibinfo {author} {\bibfnamefont {R.}~\bibnamefont
  {Kappl}},\ }\href {https://doi.org/10.1016/j.cpc.2016.05.025} {\bibfield
  {journal} {\bibinfo  {journal} {Computer Physics Communications}\ }\textbf
  {\bibinfo {volume} {207}},\ \bibinfo {pages} {386} (\bibinfo {year}
  {2016})},\ \Eprint {https://arxiv.org/abs/1511.07875} {arxiv:1511.07875
  [astro-ph.SR]} \BibitemShut {NoStop}%
\end{thebibliography}%

\end{document}